\documentclass[a4paper,11pt,oneside]{article}
\usepackage{amsmath,amsfonts,amssymb}
\usepackage{mathrsfs}
\usepackage{graphicx}

\title{Stress tensor fluctuations in de Sitter spacetime}
\author{Guillem P\'erez-Nadal${}^1$, Albert Roura${}^2$ and Enric Verdaguer${}^1$}
\date{~}

\begin{document}

\maketitle

	\vskip -0.8cm
    {\centerline{\it ${}^{1}$~Departament de F\' isica Fonamental and Institut de Ci\`encies del Cosmos}}
    {\centerline{\it Universitat de Barcelona}}
    {\centerline{\it Av. Diagonal 647, 08028 Barcelona, Spain}}
    \vskip 0.6cm
    {\centerline{\it ${}^{2}$~Max-Planck-Institut f\"ur Gravitationsphysik}}
    {\centerline{\it (Albert-Einstein-Institut)}}
    {\centerline{\it Am M\"uhlenberg 1, 14476 Golm, Germany}}
    \vskip 1cm

\begin{abstract}
The two-point function of the stress tensor operator of a quantum field in de Sitter spacetime is calculated for an arbitrary number of dimensions. We assume the field to be in the Bunch-Davies vacuum, and formulate our calculation in terms of de Sitter-invariant bitensors. Explicit results for free minimally coupled scalar fields with arbitrary mass are provided. We find long-range stress tensor correlations for sufficiently light fields (with mass $m$ much smaller than the Hubble scale $H$), namely, the two-point function decays at large separations like an inverse power of the physical distance with an exponent proportional to $m^2/H^2$. In contrast, we show that for the massless case it decays at large separations like the fourth power of the physical distance. There is thus a discontinuity in the massless limit.
As a byproduct of our work, we present a novel and simple geometric interpretation of de Sitter-invariant bitensors for pairs of points which cannot be connected by geodesics.
\end{abstract}

\section{Introduction}
\label{sec:introduction}

Two key elements in our present understanding of cosmology \cite{mukhanov05,weinberg08} are the existence of an early inflationary phase (which successfully explains the an\-iso\-tropies of the cosmic microwave background and the observed large-scale structure of the cosmos) and the present accelerated expansion of the universe, which could be entirely driven by a non-vanishing cosmological constant according to present observations. The geometry of most inflationary models is close to de Sitter spacetime and so will be the late-time behavior of our expanding universe if it is currently driven by a cosmological constant. Therefore, the physics of de Sitter spacetime may be crucial to understanding the very early universe as well as its ultimate fate.

The standard theoretical analysis relies on the linearized
calculation of primordial cosmological perturbations generated in inflationary models \cite{mukhanov92}, and the absence of large quantum corrections to the background dynamics of a spacetime driven by a cosmological constant. However, a growing number of recent studies \cite{mukhanov97,abramo97,abramo99,losic05,losic08} claim that nonlinear contributions from cosmological perturbations in inflationary models could actually be important.
It has also been argued that higher-order radiative corrections from graviton loops could give rise to a secular screening of the cosmological constant 
for pure gravity in de Sitter spacetime \cite{tsamis96a,tsamis97}.
Such nonlinear effects are dominated by infrared modes (copiously generated by the exponential expansion) and should be fairly insensitive to the short-distance behavior of a full theory of quantum gravity. However, when quantizing metric perturbations and considering their loop contributions, especial care is needed to make sure, by introducing appropriate diffeomorphism-invariant observables, that any nontrivial results obtained do not correspond to gauge artifacts \cite{unruh98,abramo02a,geshnizjani02,geshnizjani05,tsamis05,garriga08}.

Our interest centers on the quantum metric fluctuations around exact de Sitter spacetime. At the linear level, these have been studied by several authors \cite{allen87a,antoniadis91,hawking00b,higuchi03}. The two-point function of the linearized Riemann tensor $R_{ab}^{\ \ cd}$ (which is gauge-invariant around any maximally symmetric space) can be read from the results of Ref.~\cite{kouris01}, and it decays like the fourth power of the physical distance between the two points. Here we would like to go beyond the linear level and study the effects of matter loops on the quantum metric fluctuations. A key ingredient in this respect is the two-point function of the stress tensor operators of the matter fields. Calculating this object in de Sitter spacetime, and in a de Sitter-invariant vacuum, will be the main goal of this paper. This will be done using the formalism of de Sitter-invariant bitensors \cite{allen86}. Besides offering a powerful computational technique, within this approach the results are directly obtained in a manifestly de Sitter-invariant form. This is particularly convenient, since one of the aims of this line of research is to gain insight on the possibility of a breaking of de Sitter invariance due to infrared quantum effects, as suggested by Tsamis and Woodard \cite{tsamis96a,tsamis97} or Polyakov \cite{polyakov08} (see also Refs.~\cite{alvarez09,akhmedov09} for related work concerning the latter proposal).

Explicit results for free scalar fields in the Bunch-Davies vacuum \cite{bunch78a,birrell94} are obtained. We assume minimal coupling to the curvature, but the mass $m$ of the field is arbitrary. We find that, when $m$ is small, the correlations have a long range: the stress tensor two-point function decays at large separations like an inverse power of the physical distance, with an exponent proportional, in the limit of small masses, to $m^2/H^2$, where $H$ is the Hubble constant. Hence, the correlations can decay extremely slowly for sufficiently light fields.
The strictly massless case is more subtle. As is well-known, in this case the Bunch-Davies vacuum does not exist, and in fact there is no de Sitter-invariant vacuum at all \cite{allen85}. Nevertheless, as explained in the appendix, one can consider a vacuum ``arbitrarily close'' to being de Sitter-invariant. We calculate the stress tensor two-point function of the massless field in this vacuum, and obtain a de Sitter-invariant result. Interestingly, we find that in this case there are no long-range stress tensor correlations, which instead decay at large separations like the fourth power of the physical distance. Such a discontinuity of the massless limit may seem surprising at first sight, but in fact it can be understood in simple terms. The key point is that in the massive case there is a contribution from the mass term in the stress tensor that does not vanish in the limit $m\to 0$, due to the infrared divergence of the Bunch-Davies vacuum.

Finally, we would like to highlight a simple geometric interpretation of de Sitter-invariant bitensors for pairs of points that cannot be connected by geodesics which is presented in Sec.~\ref{subsec:disconnected} and, to the best of our knowledge, has not been pointed out before. The basic de Sitter-invariant bitensors, in terms of which all other invariant bitensors can be constructed, are the geodesic distance between two points (i.e., the  length of the shortest geodesic that joins them) and its first two covariant derivatives, which also have a simple geometric interpretation. However, in de Sitter spacetime there are pairs of points that cannot be connected by a geodesic. In Ref.~\cite{allen86} it was shown that, for these pairs of points, one can still define the geodesic distance and its covariant derivatives by analytic continuation, but no geometric interpretation of the resulting objects was provided. We give this interpretation in Sec.~\ref{subsec:disconnected}. 

The paper is organized as follows.
We start by reviewing in Sec.~\ref{sec:deSitter} the basic geometric properties of de Sitter spacetime that will be used in the rest of the paper (including a simple proof of the existence of pairs of points that cannot be connected by geodesics). The definition of bitensors and their general properties are introduced in Sec.~\ref{sec:bitensors} as well as those of maximally symmetric bitensors. The latter are studied in more detail for the particular case of de Sitter spacetime in Sec.~\ref{sec:dS_invariant}, and a novel simple geometric interpretation for geodesically disconnected points is presented. These tools are employed in Sec.~\ref{sec:st_fluctuations} to write the stress tensor two-point function for spacelike separated points in de Sitter-invariant form, and explicit expressions for the case of a free minimally coupled scalar field are derived. Furthermore, closed results in terms of elementary functions are given in Sec.~\ref{sec:particular} for the particular cases of small masses, vanishing mass, and large separations. The extension to timelike separated points is provided in Sec.~\ref{sec:timelike}. We discuss the implications of our results for the quantum metric fluctuations around de Sitter spacetime in Sec.~\ref{sec:discussion}. The subtleties concerning the vacua of massless fields in the case of minimal coupling are discussed in the appendix.

We use the $(+,+,+)$ sign convention of Ref.~\cite{misner73} and the abstract index notation of Ref.~\cite{wald84}. Throughout the paper we work with natural units ($\hbar=c=1$) such that the Hubble constant  equals one, unless explicitly stated otherwise.

\section{De Sitter spacetime}
\label{sec:deSitter}

\begin{figure}
\centering
\includegraphics[scale=1]{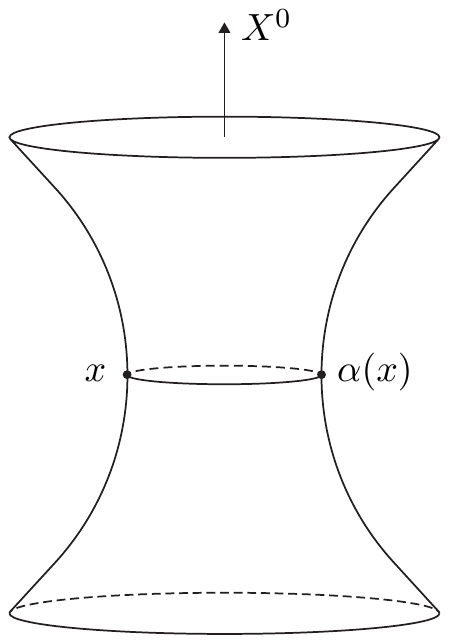}
\caption{Hyperboloid representing de Sitter spacetime. The point $x$ and its antipode, $\alpha(x)$, are also represented, as well as a spacelike geodesic joining them.}
\label{hyp}
\end{figure}

We start by recalling the definition of de Sitter spacetime. Consider the $(D+1)$-dimensional Minkowski spacetime, ${\cal M}^{D+1}$, with metric $\eta_{AB}$. The $D$-dimensional de Sitter spacetime, $dS_{D}$, is the set of position vectors $X^A\in{\cal M}^{D+1}$ that satisfy the equation
\begin{equation}
\eta_{AB}X^AX^B= H^{-2},
\label{dS}
\end{equation}
where $H$ is a parameter with dimensions of mass called the Hubble constant. This equation defines a hyperboloid, represented in Fig.~\ref{hyp}.
From this definition we see that $dS_D$ is the set of points of ${\cal M}^{D+1}$ that lie at a constant distance from the origin, and thus it is the Minkowskian analog of the sphere, $S^D$.

The isometry group of $dS_D$ is formed by all the isometries of ${\cal M}^{D+1}$ that leave the hyperboloid (\ref{dS}) invariant. These are the Lorentz transformations. Thus, the isometry group of $dS_D$ is the Lorentz group, $O(D,1)$. Since $O(D,1)$ is generated by $D(D+1)/2$ Killing vectors, which is the maximum number of Killing vectors that a manifold can have, $dS_D$ is a maximally symmetric space, like $S^D$. The group $O(D,1)$ is also called the de Sitter group.


It can be shown that the metric of $dS_D$, $g_{ab}$, is a solution of Einstein's equations,
\begin{equation}
G_{ab}+\Lambda g_{ab}=0,
\end{equation}
where $G_{ab}$ is the Einstein tensor, and
\begin{equation}
\Lambda=\frac{(D-1)(D-2)}{2}H^2
\end{equation}
is the cosmological constant. Henceforth we will set $H=1$. The dependence on $H$ can be restored throughout by dimensional analysis.

\subsection{Points not connected by geodesics}

In $dS_D$, some spacelike separated points cannot be connected by a geodesic. In order to see this, let us first introduce the biscalar 
\begin{equation}
Z(x,x')=\eta_{AB}X^A(x)X^B(x'),
\label{z}
\end{equation}
where $x,x'\in dS_D$. Note that $Z$ is invariant under the de Sitter group: if $\sigma$ is an isometry, i.e., $X^A(\sigma(x))=\Lambda^{A}_{\ B}X^B(x)$, where $\Lambda^{A}_{\ B}$ is a Lorentz transformation, then $Z(\sigma(x),\sigma(x'))=Z(x,x')$. It is also useful to introduce the Minkowskian distance, $d(x,x')$, defined by
\begin{equation}
d^2(x,x')=\eta_{AB}[X^A(x)-X^A(x')][X^B(x)-X^B(x')].
\label{d}
\end{equation}
That is, $d(x,x')$ is the length of the straight line in the embedding ${\cal M}^{D+1}$ that joins $x$ and $x'$. $Z$ and $d$ are related through
\begin{equation}
d^2=2(1-Z),
\label{dz}
\end{equation}
as follows from Eq.~(\ref{dS}). Therefore, $Z=1$ when $x'$ is on the light cone of $x$, $Z>1$ when the points are timelike separated, and $Z<1$ when the points are spacelike separated. It should also be noted that, as the squared Minkowskian distance between two points on the hyperboloid can take any value, $Z\in (-\infty,\infty)$. 

Consider now the spacelike curve represented in Fig.~\ref{hyp}, passing through the point $x$. This curve is a geodesic, because it is a straight line on a cylinder tangent to the hyperboloid. The Minkowskian distance between $x$ and any other point $x'$ in this geodesic satisfies $0\leq d(x,x')\leq 2$. This implies 
\begin{equation}
-1\leq Z(x,x')\leq 1. 
\label{bou}
\end{equation}
Now, any other spacelike geodesic passing through $x$ can be obtained from this one by an isometry that leaves $x$ invariant. This is analogous to the case of the sphere, where any geodesic (meridian) passing through the north pole is obtained from a given one by a rotation that leaves the north pole invariant. This means that any point $y'$ connected to $x$ by a spacelike geodesic satisfies $y'=\sigma(x')$ for some $x'$ in the geodesic represented in Fig.~\ref{hyp}, and where $\sigma$ is an isometry such that  $\sigma(x)=x$. De Sitter invariance of $Z$ then implies $Z(x,y')=Z(\sigma(x),\sigma(x'))=Z(x,x')$. Consequently, equation (\ref{bou}) holds for any point $x'$ connected with $x$ by a spacelike geodesic. In other words, spacelike separated points $x,x'$ such that $Z(x,x')<-1$ cannot be joined by a geodesic. 

\begin{figure}
\centering
\includegraphics[scale=1]{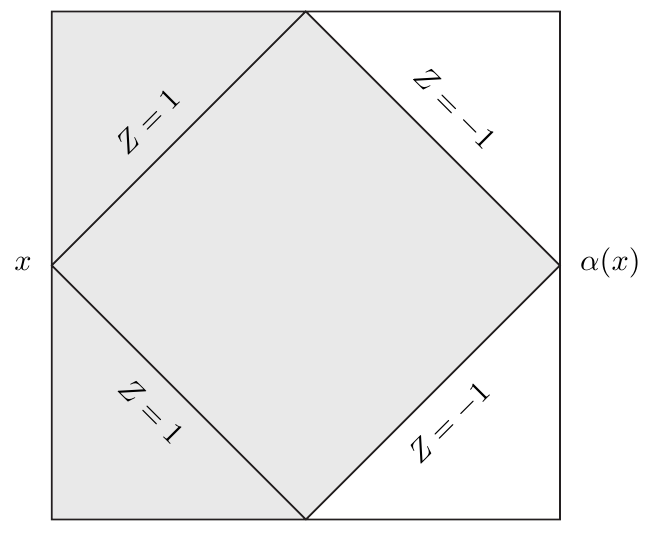}
\caption{The conformal diagram of de Sitter spacetime. The points in the shaded regions are geodesically connected with $x$.}
\label{conf}
\end{figure}

There is no similar restriction for timelike geodesics, as there are theorems that guarantee the existence of geodesics between any two timelike separated points when the spacetime is globally hyperbolic \cite{wald84}. The regions of $dS_D$ that are geodesically connected to a point $x$ are depicted in the conformal diagram of $dS_D$ in Fig.~\ref{conf}. 

\subsection{The antipodal transformation}

We will devote special attention to one particular isometry of $dS_D$ called antipodal transformation, that we denote by $\alpha$. This operation sends a point $x$ to its antipodal point $\alpha(x)$, defined by $X^A(\alpha(x))=-X^A(x)$. An example is given in Fig.~\ref{hyp}. This isometry has some special features, namely that it commutes with the rest of elements of the de Sitter group, $\sigma(\alpha(x))=\alpha(\sigma(x))$,
and that its square is the identity, $\alpha(\alpha(x))=x$. On the other hand, it follows directly from the definition of the antipodal transformation that 
\begin{equation}
Z(\alpha(x), x')=-Z(x,x').
\end{equation}
Consequently, when $x'$ is on the light cone of $\alpha(x)$, then $Z(x,x')=-1$. This is shown in Fig.~\ref{conf}.

\section{Bitensors}
\label{sec:bitensors}

In this paper we want to compute the stress tensor two-point function of a quantum field. What kind of object is this, from a geometric point of view? It is not a tensor, because it takes values on every pair of points of the spacetime, and it has indices associated with each of the points. It is rather a bitensor. Bitensors were first introduced by Synge \cite{synge60}. In this section we will give a precise definition of what bitensor means, and summarize some results that will be important for us.

Let $M$ be a manifold. Consider two points $x, x'\in M$, and let $V_x$ and $V_{x'}$ be the tangent spaces at $x$ and $x'$ respectively. A bitensor, $T_{a\dots b'}$, of type $(0,k)(0,k')$ over $V_x$ and $V_{x'}$ is a multilinear application\footnote{One can define more general bitensors, of type $(j,k)(j',k')$, which have superindices as well as subindices. Here we restrict ourselves to bitensors of type $(0,k)(0,k')$ for simplicity.}
\begin{equation*}
T_{a\dots b'}:V_x\times\dots\times V_x\times V_{x'}\times\dots\times V_{x'}\to {\mathbb C}
\end{equation*}
That is, given $k$ vectors at the point $x$ and $k'$ vectors at the point $x'$, $T_{a\dots b'}$ produces a complex number. As we see, the definition of a bitensor resembles that of an ordinary tensor. In particular, a real-valued bitensor of type $(0,k)(0,0)$ is a tensor of type $(0,k)$. 

A bitensor field, $T_{a\dots b'}$, is a rule that assigns a bitensor over $V_x$ and $V_{x'}$, $T_{a\dots b'}(x,x')$, to each pair of points $(x,x')\in M\times M$. Note that a real-valued bitensor field of type $(0,k)(0,0)$ is not a tensor field, but a rule that assigns a tensor over $V_x$ to each pair of points $(x,x')$. In spite of this, we will refer to bitensor fields simply as bitensors, as is commonly done with ordinary tensor fields.

A bitensor can be differentiated with respect to either of its arguments. The covariant derivative of a bitensor, $T_{a\dots b'}$, of type $(0,k)(0,k')$ with respect to $x$, which is denoted by $\nabla_{c}T_{a\dots b'}$, is obtained by (i) fixing the value of $x'$; and (ii) taking the covariant derivative of the resulting object as if it were a tensor of type $(0,k)$. The covariant derivative with respect to $x'$ is defined analogously. 

Bitensors of type $(0,0)(0,0)$ and $(0,1)(0,1)$ are also called biscalars and bivectors respectively. Bitensors are central mathematical objects in quantum field theory, because any two-point correlation function is a bitensor. For example, the propagators of the scalar, the photon and the graviton fields are a biscalar, a bivector and a bitensor of type $(0,2)(0,2)$ respectively.

\subsection{Invariance of a bitensor}

The notion of invariance of a bitensor is, again, closely related to the well-known notion of invariance of a tensor. Consider a diffeomorphism $\sigma:M\to M$, that assigns a point $\sigma(x)$ to each point $x$, and let $\sigma^*$ be the pushforward associated with this diffeomorphism. That is, if the vector $v^a\in V_x$ is tangent to the curve $x(\lambda)$ at the point $x$, then the vector $\sigma^*v^a\in V_{\sigma(x)}$ is tangent to the curve $\sigma(x(\lambda))$ at the point $\sigma(x)$. A bitensor $T_{a\dots b'}$ is said to be invariant under $\sigma$ if
\begin{equation}
T_{a\dots b'}(\sigma(x),\sigma(x'))\sigma^*v^a\dots \sigma^*w^{b'}=T_{a\dots b'}(x,x')v^a\dots w^{b'}
\label{inv}
\end{equation}
for all pairs of points $(x,x')$, and for any $v^a,\dots$ $\in V_x$ and $\dots, w^{b'}\in V_{x'}$. An important property concerning invariance of bitensors is that, in the case where $\sigma$ is an isometry, the covariant derivative of a bitensor invariant under $\sigma$ is also invariant under $\sigma$.

\subsection{Maximally symmetric bitensors}

Consider the case where $M$ is maximally symmetric. A bitensor is said to be maximally symmetric if it is invariant under all the isometries of $M$. The following bitensors are maximally symmetric \cite{allen86}: the distance, $\mu(x,x')$, along the shortest geodesic joining $x$ and $x'$, also called the geodesic distance; the unit vectors, $n_a(x,x')$ and $n_{a'}(x,x')$, tangent to the geodesic at the points $x$ and $x'$ respectively, pointing outward from it; the parallel propagator, $g_{ab'}(x,x')$, which parallel-transports a vector from $x'$ to $x$ along the geodesic; and the metric tensors, $g_{ab}(x,x')$ and $g_{a'b'}(x,x')$, at the points $x$ and $x'$ respectively. Furthermore, it was shown in Ref.~\cite{allen86} that these bitensors form a complete set, in the following sense: any maximally symmetric bitensor is a linear combination of products of $n_a,n_{a'},g_{ab},g_{a'b'}$ and $g_{ab'}$, with coefficients that depend only on $\mu$.

It is worth being more precise with the definition of $\mu, n_a$ and $n_{a'}$. If we denote the shortest geodesic joining $x$ and $x'$ by $x(\lambda)$ and its tangent vector by $v^a(\lambda)$, and we take $x(0)=x$ and $x(1)=x'$, the geodesic distance between $x$ and $x'$ is
\begin{equation}
\mu(x,x')=\int_{0}^{1}d\lambda\, [g_{ab}(x(\lambda))v^a(\lambda)v^b(\lambda)]^{1/2}.
\label{geo}
\end{equation}
Note that $\mu(x,x')$ is a real number for spacelike separated points, and an imaginary number for timelike separated points. The sign prescription for the square root is chosen in such a way that $\mu(x,x')>0$ for spacelike separated points, and $\mu(x,x')=i\xi(x,x')$, with $\xi(x,x')>0$, for timelike separated points. The bitensors $n_a$ and $n_{a'}$ are defined by the equations
\begin{equation}
n_a=\nabla_a\mu\qquad n_{a'}=\nabla_{a'}\mu.
\label{geovec}
\end{equation}
The geometric interpretation of $n_a(x,x')$ and $n_{a'}(x,x')$ is obtained by differentiating Eq.~(\ref{geo}). One finds that they are the unit vectors tangent to the geodesic at the points $x$ and $x'$, pointing outward from it, and multiplied by the imaginary unit in the case where $x$ and $x'$ are timelike separated. This implies $n_an^a=n_{a'}n^{a'}=1$ for both spacelike and timelike separated points. 

Some useful properties of $n_a, n_{a'}$ and $g_{ab'}$, which follow directly from the definition of these bitensors, are
\begin{eqnarray}
&&g_{ab'}(x,x')n^{b'}(x,x')=-n_a(x,x')\nonumber\\
&&g_{a}^{\ c'}(x,x')g_{c'b}(x',x)=g_{ab}(x).
\end{eqnarray}
From the last of these equations, and the fact that parallel transport leaves the scalar product invariant, one can prove the additional relation
\begin{equation}
g_{ab'}(x,x')=g_{b'a}(x',x).
\end{equation}
The theorem of Ref.~\cite{allen86} can be applied to the computation of the covariant derivatives of $n_a, n_{a'}$ and $g_{ab'}$, which are also maximally symmetric bitensors. The corresponding $\mu$-dependent coefficients were also computed in Ref.~\cite{allen86} for all maximally symmetric spaces. In the particular case of $dS_D$, the result is 
\begin{eqnarray}
&&\nabla_{a}n_b=\cot\mu\,(g_{ab}-n_an_b)\nonumber\\
&&\nabla_{a}n_{b'}=-\csc\mu\,(g_{ab'}+n_an_{b'})\nonumber\\
&&\nabla_a g_{bc'}=(\csc\mu-\cot\mu)(g_{ab}n_{c'}+g_{ac'}n_b).
\label{cov}
\end{eqnarray}

\section{De Sitter-invariant bitensors}
\label{sec:dS_invariant}

The bitensors $\mu, n_a, n_{a'}$ and $g_{ab'}$ of $dS_D$ are defined in the domain
\begin{equation*}
{\cal G}=\{(x,x')\in dS_D\times dS_D{\text{ such that $\exists$ a geodesic joining $x$ and $x'$}}\}.
\end{equation*}
As we know from Sec.~\ref{sec:deSitter}, this does not cover the whole $dS_D\times dS_D$: only the pairs of points with $Z\ge -1$ belong to ${\cal G}$, where $Z$ has been defined in Eq.~(\ref{z}). However, as was shown in Ref.~\cite{allen86}, the definition of $\mu, n_a, n_{a'}$ and $g_{ab'}$ can be extended to pairs of points outside ${\cal G}$ in such a way that all their properties are preserved. Let us review the argument of Ref.~\cite{allen86}.

\begin{figure}
\centering
\includegraphics[scale=1]{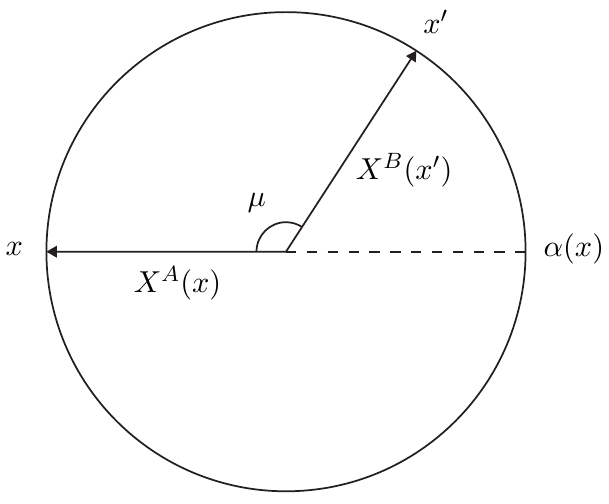}
\caption{The geodesic of Fig.~\ref{hyp}, seen from above. The geodesic distance $\mu(x,x')$ between $x$ and $x'$ is the angle between the position vectors $X^A(x)$ and $X^B(x')$.}
\label{figzmu}
\end{figure}

For spacelike geodesics, $-1\le Z< 1$, the geodesic distance $\mu$ of $dS_D$ is given by the equation
\begin{equation}
\mu=\cos^{-1}Z,
\label{muz}
\end{equation}
where $\cos^{-1}$ is the principal value of the inverse cosine, which is single-valued. For example, $\cos^{-1}1=0$, and $\cos^{-1}(-1)=\pi$. For the geodesic of Fig.~\ref{hyp}, this equation is easily seen from Fig.~\ref{figzmu}.
Other spacelike geodesics are obtained from this one by an isometry, and therefore, by de Sitter invariance of $\mu$ and $Z$, Eq.~(\ref{muz}) is true for all spacelike geodesics. 

Using a similar argument, it can be shown that a similar equation holds for timelike separated points, $Z>1$:
\begin{equation}
\mu=\cos^{-1}(Z-i\epsilon),
\label{muz2}
\end{equation}
where $\epsilon$ is a positive real infinitesimal. This $-i\epsilon$ term is added because the function $\cos^{-1}Z$ has a branch cut at $Z<-1$ and $Z>1$ along the real axis, across which its imaginary part changes sign.

Eqs.~(\ref{muz}) and (\ref{muz2}) serve to define $\mu$ outside ${\cal G}$ by analytic continuation. Let us label the points of $dS_D$ with their corresponding position vectors in the embedding ${\cal M}^{D+1}$. The geodesic distance, $\mu(X^A,Y^A)$, is only defined for pairs of points $(X^A,Y^A)$ such that $Z(X^A,Y^A)\ge-1$. However, $Z(X^A,Y^A)$ is an analytic (polynomial) function of $(X^A,Y^A)$ for all complex values of $X^A$ and $Y^A$. Furthermore, $\cos^{-1}Z$ is an analytic function of $Z$ in the whole complex plane except the branch cut. Therefore, $\cos^{-1}Z$ is the analytic continuation of $\mu$ to complex values of $X^A$ and $Y^A$. We can define $\mu$ for pairs of points with $Z<-1$ as
\begin{equation}
\mu=\cos^{-1}(Z+i\epsilon),
\label{muz3}
\end{equation}
where the sign of the $i\epsilon$ term has been taken positive for convenience. One could also adopt the opposite choice, in which case the imaginary part of the geodesic distance $\mu$ outside ${\cal G}$ would have the opposite sign.

The definition of $n_a$ and $n_{a'}$ is extended outside ${\cal G}$ by substituting (\ref{muz3}) in (\ref{geovec}). The definition of the parallel propagator can also be extended using the second equation in (\ref{cov}), which gives $g_{ab'}$ in terms of $\mu, n_a$ and $n_{a'}$. Since these definitions have been obtained by analytic continuation, all the results concerning $\mu,n_a,n_{a'}$ and $g_{ab'}$ that we have presented in the previous section hold everywhere in $dS_D\times dS_D$.

\subsection{Geometric interpretation}
\label{subsec:disconnected}

What was not emphasized in Ref.~\cite{allen86} is that the bitensors $\mu,n_a,n_{a'}$ and $g_{ab'}$ also have a geometric interpretation outside ${\cal G}$. We will explain it here. The principal value of the inverse cosine has the property
\begin{equation}
\cos^{-1}Z=\pi-\cos^{-1}(-Z).
\end{equation}
On the other hand, as we know from Sec.~\ref{sec:deSitter}, $Z(\alpha(x),x')=-Z(x,x')$. Therefore, the equation above and (\ref{muz})-(\ref{muz3}) imply 
\begin{equation}
\mu(x,x')=\pi-\mu(\alpha(x),x')
\label{teo}
\end{equation}
throughout $dS_D\times dS_D$. For pairs of points in the geodesic of Fig.~\ref{hyp}, this can also be seen from Fig.~\ref{figzmu}. Now, for $(x,x')\notin {\cal G}$ we have $Z(x,x')<-1$, which implies $Z((\alpha(x),x')>1$. Therefore, $\alpha(x)$ and $x'$ are geodesically connected and timelike separated. This can also be seen in Fig.~\ref{conf}. This means that the geodesic distance for geodesically disconnected points is $\pi$ minus the length of the shortest geodesic joining $\alpha(x)$ and $x'$. Since this geodesic is timelike, the geodesic distance is a complex number outside ${\cal G}$: $\mu=\pi-i\xi$, with $\xi\in {\mathbb R}^+$.

\begin{figure}
\centering
\includegraphics[scale=1]{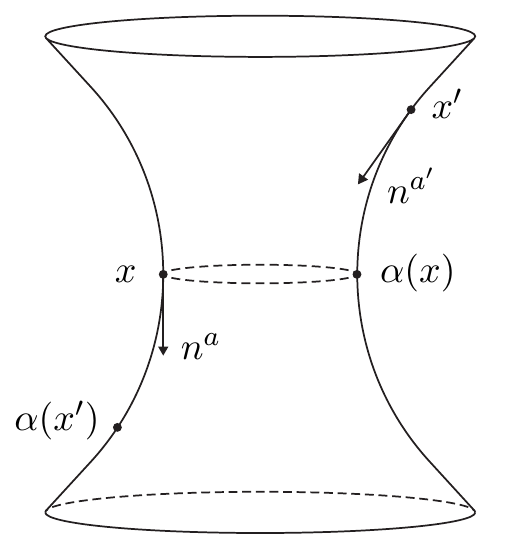}
\caption{Example of unit vectors $n_a$ and $n_{a'}$ for points $x$ and $x'$ that are not connected by a geodesic.}
\label{figvec}
\end{figure}

Substituting (\ref{teo}) in (\ref{geovec}), we obtain
\begin{equation}
n_a(x,x')=-n_a(x,\alpha(x'))\qquad n_{a'}(x,x')=-n_{a'}(\alpha(x),x'),
\label{teo2}
\end{equation}
where we have used that $\mu(\alpha(x),x')=\mu(x,\alpha(x'))$. This is because $\mu$ is de Sitter-invariant, and the antipodal transformation is an isometry such that $\alpha(\alpha(x))=x$. Eq.~(\ref{teo2}) means that, for $(x,x')\notin {\cal G}$, $n_a(x,x')$ is the unit vector tangent at $x$ to the geodesic joining $x$ and $\alpha(x')$, and $n_{a'}(x,x')$ is the unit vector tangent at $x'$ to the geodesic joining $\alpha(x)$ and $x'$, both of them pointing inward. This is shown in Fig.~\ref{figvec}.

Finally, after substitution of (\ref{teo}) and (\ref{teo2}) in the second equality of (\ref{cov}), we obtain for the parallel propagator
\begin{equation}
g_{ab'}=-\bar g_{ab'}-2n_an_{a'},
\label{teo3}
\end{equation}
where the bivector $\bar g_{ab'}$ is defined by the equation
\begin{equation}
\bar g_{ab'}(x,x')w^{b'}=g_{ab'}(x,\alpha(x'))\alpha^*w^{b'}
\end{equation}
for all vectors $w^{b'}\in V_{x'}$. In order to obtain Eq.~(\ref{teo3}), we have used the property $n_{b'}(\alpha(x),x')w^{b'}=n_{b'}(x,\alpha(x'))\alpha^*w^{b'}$, which holds for all vectors $w^{b'}\in V_{x'}$, and which is a direct consequence of de Sitter invariance of $n_{b'}$ and the properties of the antipodal transformation. For $(x,x')\notin {\cal G}$, $\bar g_{ab'}(x,x')$  is the operator that ``pushes forward'' a vector from $x'$ to $\alpha(x')$, and then parallel transports it from $\alpha(x')$ to $x$ along the geodesic joining these points.

\subsection{A comment on the Minkowskian distance}
\label{subsecmin}

In the previous subsection we have found the geometric interpretation of the bitensors $\mu,n_a,n_{a'}$ and $g_{ab'}$ for pairs of points that are not connected by a geodesic. In the particular case of the geodesic distance $\mu$, note that, although the geometric interpretation is simple, $\mu$ is not the length of any spacelike curve (remember that $x$ and $x'$ are spacelike separated when they are not geodesically connected). Therefore, it cannot be taken as a notion of distance. In order to interpret physically our results below for spacelike separated points, it will be more convenient to express them in terms of the Minkowskian distance, defined in Eq.~(\ref{d}). The reason is that, for $x$ and $x'$ spacelike separated, this biscalar coincides with the so-called ``physical distance'' or ``proper distance'' associated with the flat slicing of $dS_D$. Let us see this. If we label the points in $dS_D$ with the coordinates $\{t,x^i\}$, with $i=1,\dots, D-1$, defined by
\begin{eqnarray}
&&X^0=-\sinh t-\frac{1}{2}\delta_{ij}x^ix^je^t\nonumber\\
&&X^i=x^i e^t\nonumber\\
&&X^D=\cosh t-\frac{1}{2}\delta_{ij}x^ix^je^t,
\label{flatco}
\end{eqnarray}
then the line element takes the form
\begin{equation}
ds^2=\eta_{AB}dX^AdX^B=-dt^2+e^{2t}\delta_{ij}dx^idx^j,
\end{equation}
so that the surfaces $\Sigma_t$ of constant $t$ are flat. Moreover, any two spacelike separated points can be brought to the same flat hypersurface by an isometry. Now, for $x,x'\in \Sigma_t$ we have
\begin{equation}
d^2(x,x')=e^{2t}\delta_{ij}[x^i(x)-x^i(x')][x^j(x)-x^j(x')],
\end{equation}
as can be straightforwardly seen by using Eqs.~(\ref{d}) and (\ref{flatco}). Hence, for $x,x'\in \Sigma_t$ the Minkowskian distance coincides with the length of the straight line of $\Sigma_t$ that joins $x$ and $x'$. This is what in cosmology is called ``physical distance'' or ``proper distance''.

\section{Stress tensor fluctuations in de Sitter spacetime}
\label{sec:st_fluctuations}

Consider a quantum field in $dS_D$, whose stress tensor operator is $T_{ab}$, in a vacuum state $|0\rangle$ that we assume de Sitter-invariant. We want to compute the two-point function of the stress tensor operator,
\begin{equation}
F_{abc'd'}(x,x')=\langle 0| T_{ab}(x)T_{c'd'}(x') |0\rangle-\langle 0| T_{ab}(x) |0\rangle \langle 0| T_{c'd'}(x') |0\rangle,
\label{flu}
\end{equation}
which we will also refer to as the stress tensor fluctuations. For the moment we will assume that the points $x$ and $x'$ are spacelike separated. What we have seen so far tells us a lot about this two-point function before doing any computation, even before specifying the quantum field theory under consideration. Indeed, if we choose $|0\rangle$ to be a de Sitter-invariant state, the expectation value of any observable must be de Sitter-invariant. If the points $x$ and $x'$ are spacelike separated, causality requires
\begin{equation}
[T_{ab}(x),T_{c'd'}(x')]=0,
\label{com1}
\end{equation}
which implies that $F_{abc'd'}$ is the expectation value of a Hermitian operator, i.e., an observable (the product of two Hermitian operators that commute is Hermitian). Therefore, when $x$ and $x'$ are spacelike separated, $F_{abc'd'}$ is a de Sitter-invariant bitensor. According to the theorem stated in the previous section, it must be a linear combination of products of $n_a,n_{a'},g_{ab},g_{a'b'}$ and $g_{ab'}$, with coefficients that depend only on $\mu$.
On the other hand, for spacelike separated points Eq.~(\ref{com1}) implies 
\begin{equation*}
F_{abc'd'}(x,x')=F_{c'd'ab}(x',x).
\end{equation*}
And, since the stress tensor is symmetric, we have 
\begin{equation*}
F_{abc'd'}=F_{bac'd'}=F_{abd'c'}.
\end{equation*}
The most general bitensor of type $(0,2)(0,2)$ (i.e., with the index structure of $F_{abc'd'}$) that is a linear combination of products of $n_a,n_{a'},g_{ab},g_{a'b'}$ and $g_{ab'}$, with coefficients that depend only on $\mu$, and that satisfies the above constraints is
\begin{eqnarray}
F_{abc'd'}&=&P(\mu)n_an_bn_{c'}n_{d'}+Q(\mu)(n_an_bg_{c'd'}+n_{c'}n_{d'}g_{ab})\nonumber\\
&&+R(\mu)(n_an_{c'}g_{bd'}+n_bn_{d'}g_{ac'}+n_an_{d'}g_{bc'}+n_bn_{c'}g_{ad'})\nonumber\\
&&+S(\mu)(g_{ac'}g_{bd'}+g_{bc'}g_{ad'})+T(\mu)g_{ab}g_{c'd'}.
\label{flu1}
\end{eqnarray}
Moreover, the stress tensor conservation law implies $\nabla^a F_{abc'd'}=0$. Imposing this equation on (\ref{flu1}), and using (\ref{cov}), we obtain
\begin{eqnarray}
P'+Q'-2R'+(D-1)P\cot\mu-2Q\csc\mu &&\nonumber\\
-2R[(D-2)\cot\mu -D\csc\mu]&=&0\nonumber\\
Q'+T'+(D-1)Q\cot\mu-2R\csc\mu-2S(\cot\mu-\csc\mu)&=&0\nonumber\\
R'-S'-Q\csc\mu+DR\cot\mu-DS(\cot\mu-\csc\mu)&=&0,
\label{con}
\end{eqnarray}
where the prime stands for derivative with respect to $\mu$. Eqs.~(\ref{flu1}) and (\ref{con}) are valid for any quantum field theory in $dS_D$, assuming that $|0\rangle$ is de Sitter-invariant and that the points are spacelike separated.

There is yet another constraint on the functions $P,Q,R,S$ and $T$, which comes from requiring the stress tensor two-point function to be unambiguous for antipodal points, $x'=\alpha(x)$. The unit vector $n_a$ is not well-defined when $x'=\alpha(x)$, because there is an infinite number of geodesics joining $x$ and $\alpha(x)$, all of them with the same length, $\mu(x,\alpha(x))=\pi$. This is analogous to what happens in the sphere: the north and the south poles are joined by an infinite number of meridians, all of them with the same length. The parallel propagator is also not well-defined, but it can be split into an ambiguous and an unambiguous part through Eq.~(\ref{teo3}). Using this equation, (\ref{flu1}) can be rewritten as
\begin{eqnarray} 
F_{abc'd'}&=&(P+8S-8R)n_an_bn_{c'}n_{d'}+Q(n_an_bg_{c'd'}+n_{c'}n_{d'}g_{ab})\nonumber\\
&&+(2S-R)(n_an_{c'}\bar g_{bd'}+n_bn_{d'}\bar g_{ac'}+n_an_{d'}\bar g_{bc'}+n_bn_{c'}\bar g_{ad'})\nonumber\\
&&+S(\bar g_{ac'}\bar g_{bd'}+\bar g_{bc'}\bar g_{ad'})+Tg_{ab}g_{c'd'}.
\end{eqnarray}
The bitensors appearing in this equation are all well-defined for antipodal points except for $n_a$. In order for the stress tensor two-point function to be unambiguous when $x'=\alpha(x)$, the coefficients of the terms containing $n_a$ in the equation above should vanish at $\mu=\pi$,
\begin{equation}
P(\pi)+8S(\pi)-8R(\pi)=Q(\pi)=2S(\pi)-R(\pi)=0.
\label{ant}
\end{equation}
This equation is also valid for any quantum field theory in $dS_D$, assuming that $|0\rangle$ is de Sitter-invariant.

\subsection{Stress tensor fluctuations of a free scalar field}
\label{subsecfree}

Let us be more concrete and consider a particular quantum field theory in $dS_D$, namely that of a free, minimally coupled scalar field $\phi$, with mass $m$. The corresponding equation of motion is the Klein-Gordon equation
\begin{equation}
(\nabla_a\nabla^a-m^2)\phi=0.
\label{kg}
\end{equation}
The stress tensor operator of this field can be written using point splitting regularization as 
\begin{equation}
T_{ab}(x)=\lim_{x''\to x} {\cal D}_{ab}(x,x'')\phi(x)\phi(x''),
\label{str}
\end{equation}
with
\begin{equation}
{\cal D}_{ab}=g_{b}^{\ c''}\nabla_a\nabla_{c''}-\frac{1}{2}g_{ab}g^{cd''}\nabla_c\nabla_{d''}-\frac{1}{2}m^2g_{ab}.
\label{dab}
\end{equation}
In order to check that this expression coincides with the standard one \cite{birrell94}, one has to take into account that the parallel propagator becomes the metric in the coincidence limit, i.e., $g_{ab''}(x,x'')\to g_{ab}(x)$ when $x''\to x$. The stress tensor two-point function of the free scalar field is obtained by substituting (\ref{str}) in (\ref{flu}),
\begin{eqnarray}
F_{abc'd'}(x,x')=\lim_{\substack{x''\to x\\x'''\to x'}} \!\!\!\!\!\!\!\!\!\!&&{\cal D}_{ab}(x,x''){\cal D}_{c'd'}(x',x''')\left[\langle 0|\phi(x)\phi(x'')\phi(x')\phi(x''')|0\rangle\right.\nonumber\\
&&\left.-\langle 0|\phi(x)\phi(x'')|0\rangle\langle 0|\phi(x')\phi(x''')|0\rangle\right].
\label{flu2}
\end{eqnarray}
Since the field $\phi$ is free, and $|0\rangle$ is a vacuum state, the four-point function appearing in this equation can be written in terms of the Wightman function, $G^+(x,x')=\langle 0|\phi(x)\phi(x')|0\rangle$,
\begin{eqnarray}
\langle 0|\phi(x)\phi(x'')\phi(x')\phi(x''')|0\rangle &=& G^+(x,x'')G^+(x',x''')+G^+(x,x')G^+(x'',x''')\nonumber\\
&&+G^+(x,x''')G^+(x'',x').
\label{wic}
\end{eqnarray}
This equation is obtained by expanding the field $\phi$ in terms of creation and annihilation operators. Note that, in the particular case where the product of fields on the left-hand side is time-ordered, this equation also follows from Wick's theorem. Substituting (\ref{wic}) in (\ref{flu2}), and taking the limit $x''\to x$ and $x'''\to x'$, we find
\begin{eqnarray}
F_{abc'd'}&=&\nabla_a\nabla_{c'}G^+\nabla_b\nabla_{d'}G^+ + \nabla_a\nabla_{d'}G^+\nabla_b\nabla_{c'}G^+\nonumber\\
&&-g_{ab}\left(\nabla_e\nabla_{c'}G^+\nabla^e\nabla_{d'}G^+ +m^2\nabla_{c'}G^+\nabla_{d'}G^+\right)\nonumber\\
&&-g_{c'd'}\left(\nabla_a\nabla_{e'}G^+\nabla_b\nabla^{e'}G^+ +m^2\nabla_{a}G^+\nabla_{b}G^+\right)\nonumber\\
&&+\frac{1}{2}g_{ab}g_{c'd'}\left[\nabla_e\nabla_{e'}G^+\nabla^e\nabla^{e'}G^+\right.\nonumber\\
&&\left. +m^2\left(\nabla_eG^+\nabla^eG^+ + \nabla_{e'}G^+\nabla^{e'}G^+\right)+m^4G^{+2}\right].
\label{flu3}
\end{eqnarray}
Note that, after substitution of (\ref{wic}) in (\ref{flu2}), there has been a cancellation of terms that diverge in the limit $x''\to x,x'''\to x'$, so that this result is finite as long as $x$ is not on the light cone of $x'$.

At this point we impose that the vacuum $|0\rangle$ is de Sitter-invariant, and that the points $x$ and $x'$ are spacelike separated. For spacelike separated points, the Wightman function is the expectation value of an observable, so that it must be a de Sitter-invariant biscalar. Thus it must have the form
\begin{equation}
G^+=G(\mu).
\label{wig0}
\end{equation}
On the other hand, the Wightman function satisfies the Klein-Gordon equation (\ref{kg}). Using (\ref{wig0}) and (\ref{cov}), this equation translates into
\begin{equation}
G''(\mu)+(D-1)\cot\mu G'(\mu)-m^2G(\mu)=0.
\label{kgw}
\end{equation}
Imposing as boundary conditions that $G(\mu)$ diverges only at the coincidence limit, $\mu\to 0$, and that it does so in the same way as the standard Wightman function in Minkowski spacetime, the only solution is \cite{spradlin01}
\begin{equation}
G(\mu)=c_{m,D}F(h_+,h_-;\frac{D}{2};\frac{1+ Z}{2}),
\label{wig}
\end{equation}
where $F$ denotes the hypergeometric function, $h_\pm$ and $c_{m,D}$ are constants,
\begin{eqnarray}
&&h_\pm=\frac{1}{2}\left[(D-1)\pm\sqrt{(D-1)^2-4m^2}\right]\nonumber\\
&&c_{m,D}=\frac{\Gamma(h_+)\Gamma(h_-)}{(4\pi)^{D/2}\Gamma(\frac{D}{2})},
\label{hc}
\end{eqnarray}
and we are using that $Z=\cos\mu$, as follows from (\ref{muz}) and (\ref{muz3}). One can see \cite{birrell94} that there is a vacuum whose Wightman function for spacelike separated points is (\ref{wig}). It is the so-called Bunch-Davies vacuum. In what follows we will be interested in the stress tensor fluctuations in the Bunch-Davies vacuum.

Substituting (\ref{wig0}) in (\ref{flu3}), and using (\ref{cov}), we find that the stress tensor two-point function in the Bunch-Davies vacuum has the form (\ref{flu1}), as corresponds to a de Sitter-invariant state. The associated $\mu$-dependent functions are
\begin{eqnarray}
P&=&2G_{1}^2\nonumber\\
Q&=&-G_{1}^2+2G_{1}G_2-m^2G'^2\nonumber\\
R&=&G_1G_2\nonumber\\
S&=&G_{2}^2\nonumber\\
T&=&\frac{1}{2}G_{1}^2-G_1G_2+\frac{D-4}{2}G_{2}^2+m^2G'^2+\frac{1}{2}m^4G^2,
\label{PQR}
\end{eqnarray}
where, again, the prime stands for derivative with respect to $\mu$, and
\begin{eqnarray}
G_1(\mu)&=&G''(\mu)-G'(\mu)\csc\mu\nonumber\\
G_2(\mu)&=&-G'(\mu)\csc\mu.
\label{G1G2}
\end{eqnarray}
The functions (\ref{PQR}) satisfy the stress tensor conservation equations (\ref{con}), as can be checked by using the Klein-Gordon equation (\ref{kgw}). They also satisfy the conditions (\ref{ant}), as follows from basic properties of the hypergeometric function.

\section{Particular cases}
\label{sec:particular}

Eqs.~(\ref{PQR}) give the stress tensor two-point function of the free, minimally coupled scalar field in terms of the Wightman function (\ref{wig}) and its derivatives with respect to $\mu$. Of course, this result is not very explicit. 
In this section we will derive explicit expressions for the stress tensor two-point function in the following particular cases: small masses, strictly zero mass, and large negative values of $Z$. 

The results will be given in terms of the biscalar $Z$. In order to interpret them, recall that $Z$ is related to the Minkowskian distance $d$ through Eq.~(\ref{dz}). Remember also, from Sec.~\ref{subsecmin}, that the Minkowskian distance coincides with the physical distance when the points are spacelike separated.

\subsection{Small masses}
\label{subsecsma}

Let us assume that the mass of the field is small, $m\ll 1$. Remember that we are using units in which $H=1$. In a general system of units, our assumption is that the mass is much smaller than the Hubble constant. This is the situation one encounters in inflationary models. 

We begin by studying the behavior of the Wightman function (\ref{wig}) in the limit of small masses. Using the definition of the hypergeometric function \cite{abramowitz72}, Eq.~(\ref{wig}) can be rewritten as follows for $|1+Z|/2<1$:
\begin{equation}
G(\mu)=\frac{1}{(4\pi)^{D/2}}\sum_{n=0}^{\infty}\frac{\Gamma(h_+ +n)\Gamma(h_- +n)}{\Gamma(\frac{D}{2}+n)}\frac{1}{n!}\left(\frac{1+Z}{2}\right)^n.
\label{ser}
\end{equation}
For other values of $Z$, the Wightman function is obtained by analytic continuation. The dependence on $m$ is implicit in the parameters $h_\pm$, Eq.~(\ref{hc}). 
Expanding the gamma functions in the equation above in powers of $m$ we obtain at leading order
\begin{equation}
G(\mu)=\frac{1}{m^2}\frac{\Gamma(D)}{(4\pi)^{D/2}\Gamma(\frac{D}{2})}+O(m^0).
\label{G}
\end{equation}
The Wightman function diverges in the limit $m\to 0$. This is the well-known infrared divergence of the Bunch-Davies vacuum. Since the divergent term is independent of $\mu$, the derivative of the Wightman function $G'(\mu)$ will be finite in this limit. Indeed, expanding the derivative of (\ref{ser}) in powers of the mass, we obtain after some algebraic manipulations
\begin{equation} 
G'(\mu)=-\frac{\sin\mu}{1-Z}\sum_{k=0}^{\frac{D}{2}-1}C_k\left(\frac{1+Z}{1-Z}\right)^{k}+O(m^2)
\label{Gpr}
\end{equation} 
for an even number of dimensions $D$, where
\begin{equation}
C_k=\frac{\Gamma(D/2)}{(4\pi)^{D/2}}\binom{D-1}{D/2+k}.
\end{equation}
Note that (\ref{Gpr}) is a sum over a finite number of terms, not an infinite series. Eqs.~(\ref{G}) and (\ref{Gpr}) have been derived assuming $|1+Z|/2<1$. However, since the extension to other values of $Z$ is obtained by analytic continuation, these equations are actually valid for any $Z<1$. This result generalizes to an arbitrary, even number of dimensions a previously obtained expansion of the Wightman function in powers of the mass in $dS_4$ \cite{allen87b}, and reduces to it when we set $D=4$.

Now that we have written the Wightman function and its derivative with respect to $\mu$ in terms of simple functions, we can obtain an explicit expression for the stress tensor fluctuations. Substituting (\ref{G}) and (\ref{Gpr}) in (\ref{PQR}), we find that the functions $P,Q,R,S$ have the form
\begin{equation}
F(\mu)=\frac{1}{(1-Z)^2}\sum_{n=0}^{D-2}F_n\left(\frac{1+Z}{1-Z}\right)^n+O(m^2),
\label{F}
\end{equation}
where we are denoting $P,Q,R,S$ generically by $F$. For the remaining function we obtain
\begin{equation}
T(\mu)=\frac{1}{(1-Z)^2}\sum_{n=0}^{D-2}T_n\left(\frac{1+Z}{1-Z}\right)^n+\frac{1}{2}\left[\frac{\Gamma(D)}{(4\pi)^{D/2}\Gamma(\frac{D}{2})}\right]^2+O(m^2).
\label{T}
\end{equation}
In these equations, the coefficients $F_n$ and $T_n$ are
\begin{eqnarray}
P_n&=&8\sum_{l=-N}^{N}\left(\frac{n+l}{2}+1\right)\left(\frac{n-l}{2}+1\right)C_\frac{n+l}{2}C_\frac{n-l}{2}\nonumber\\
Q_n&=&-4\sum_{l=-N}^{N}\left(\frac{n+l}{2}+1\right)\left(\frac{n-l}{2}\right)C_\frac{n+l}{2}C_\frac{n-l}{2}\nonumber\\
R_n&=&2\sum_{l=-N}^{N}\left(\frac{n+l}{2}+1\right)C_\frac{n+l}{2}C_\frac{n-l}{2}\nonumber\\
S_n&=&\sum_{l=-N}^{N}C_\frac{n+l}{2}C_\frac{n-l}{2}\nonumber\\
T_n&=&\sum_{l=-N}^{N}\left[2\left(\frac{n+l}{2}+1\right)\left(\frac{n-l}{2}\right)+\frac{D-4}{2}\right]C_\frac{n+l}{2}C_\frac{n-l}{2},
\label{PQRn}
\end{eqnarray}
where $N=\min(n,D-2-n)$. 

The stress tensor two-point function of a free scalar field with small mass, for spacelike separated points, is given by Eqs.~(\ref{F}) and (\ref{T}).
It is finite in the limit $m\to 0$, unlike the Wightman function. This is because Eqs.~(\ref{PQR}) involve only derivatives of $G$, which are finite, with the only exception of $T$, which includes a term proportional to $m^4G^2$. The factor $m^4$ cancels the divergence of the Wightman function.

An interesting aspect of the obtained stress tensor two-point function is its behavior at long Minkowskian distances. From Eq.~(\ref{dz}), this corresponds to large negative values of $Z$. Setting $Z\ll -1$ in (\ref{F}) and (\ref{T}), we obtain 
\begin{eqnarray}
P,Q,R,S &\sim & Z^{-2}+O(m^2)\nonumber\\
T&\sim &{\text{constant}}+O(m^2).
\label{lim}
\end{eqnarray}
From Eq.~(\ref{dz}), we see that the leading terms in $P,Q,R,S$ decay like $d^{-4}$. More interestingly, the leading term in $T$ does not decay to zero. In other words, {\emph{the stress tensor two-point function in the limit $m\to 0$ does not vanish at long distances}}. 

The origin of this behavior is, again, the term proportional to $m^4G^2$ in Eqs.~(\ref{PQR}). Indeed, the terms with derivatives of $G$ are either $O(m^2)$, and thus subleading, or vanishing in the limit $Z\to -\infty$, whereas the term proportional to $m^4G^2$ is a constant of order $O(m^0)$.

It is worth comparing this result with the stress tensor fluctuations in Minkowski spacetime, ${\cal M}^{D}$. Note first that, since the theorem stated in Sec.~\ref{sec:bitensors} is valid for all maximally symmetric spaces, the stress tensor fluctuations in ${\cal M}^{D}$ also have the form (\ref{flu1}). The corresponding functions $P,Q,R,S,T$ can be obtained by dimensional analysis. Indeed, these functions must have dimensions of $({\text{energy density}})^2$, that is, $({\text{mass}})^{2D}$. But the theory of a free scalar field in ${\cal M}^D$, and in the limit $m\to 0$, has no dimensionful constants. Thus, by dimensional analysis the only possibility is
\begin{equation}
P,Q,R,S,T\propto\mu^{-2D}.
\label{flumin}
\end{equation}
Hence, the stress tensor fluctuations in the limit $m\to 0$ do decay with the distance in ${\cal M}^{D}$. Finally, it is worth noting that, for small distances ($Z\to 1$) the functions (\ref{F}) and (\ref{T}) have the same form as in ${\cal M}^D$, Eq.~(\ref{flumin}), as it should be.

\subsection{Massless field}
\label{subsecmas}

We now consider a slightly different case, namely that of a strictly massless field, $m=0$. As is well-known \cite{allen85}, in this case there exists no de Sitter-invariant vacuum. The vacuum states we choose for the massless field are the $O(D)$-invariant vacua, $|0\rangle_\alpha$, introduced by Allen and Folacci \cite{allen87b}. These states form a one-parameter family with $\alpha\in (0,\infty)$. We will be interested in the limit $\alpha\to 0$, because, although $|0\rangle_\alpha$ are not de Sitter-invariant, there is a class of observables whose expectation values are de Sitter-invariant in this limit. This is explained in the appendix.

The stress tensor two-point function of a free scalar field is given by Eq.~(\ref{flu3}), for any mass $m$ and for any vacuum state. In the massless case this equation reduces to
\begin{eqnarray}
F_{abc'd'}&=&\nabla_a\nabla_{c'}G^+\nabla_b\nabla_{d'}G^+ + \nabla_a\nabla_{d'}G^+\nabla_b\nabla_{c'}G^+\nonumber\\
&&-g_{ab}\nabla_e\nabla_{c'}G^+\nabla^e\nabla_{d'}G^+-g_{c'd'}\nabla_a\nabla_{e'}G^+\nabla_b\nabla^{e'}G^+\nonumber\\
&&+\frac{1}{2}g_{ab}g_{c'd'}\nabla_e\nabla_{e'}G^+\nabla^e\nabla^{e'}G^+.
\label{flu3bis}
\end{eqnarray}
Note that all terms in this equation are products of $\nabla_a\nabla_{b'}G^{+}$. Let us denote by $F_{abc'd'}^\alpha$ the stress tensor two-point function in the vacuum $|0\rangle_\alpha$. We want to compute it in the limit $\alpha\to 0$. Now, in the appendix it is shown that
\begin{equation}
\lim_{\alpha\to 0}\nabla_a\nabla_{b'}G^{+}_\alpha=\lim_{m\to 0}\nabla_a\nabla_{b'}G^{+}_{BD},
\label{O4Bun1}
\end{equation}
where $G^{+}_\alpha$ is the Wightman function of the massless field in the vacuum $|0\rangle_\alpha$, and $G^{+}_{BD}$ is the Wightman function of the massive field in the Bunch-Davies vacuum. Therefore, the stress tensor two-point function of a massless field in the limit $\alpha\to 0$ can be computed as follows: (i) compute (\ref{flu3bis}) for a massive field in the Bunch-Davies vacuum; and (ii) let $m\to 0$. It is important to notice that this is not exactly the same as computing the stress tensor two-point function of a massive field in the limit $m\to 0$, which is what we have done in the previous subsection.

The step (i) corresponds to substitute the Wightman function (\ref{wig0}) in (\ref{flu3bis}). The result has the form (\ref{flu1}), with
\begin{eqnarray}
P&=&2G_{1}^2\nonumber\\
Q&=&-G_{1}^2+2G_{1}G_2\nonumber\\
R&=&G_1G_2\nonumber\\
S&=&G_{2}^2\nonumber\\
T&=&\frac{1}{2}G_{1}^2-G_1G_2+\frac{D-4}{2}G_{2}^2,
\label{PQR0}
\end{eqnarray}
and where $G_1$ and $G_2$ are given by (\ref{G1G2}). This equation is exactly the same as (\ref{PQR}), but without the mass terms. We can now let $m\to 0$. Following the arguments of the previous subsection, we obtain
\begin{eqnarray}
&&F(\mu)=\frac{1}{(1-Z)^2}\sum_{n=0}^{D-2}F_n\left(\frac{1+Z}{1-Z}\right)^n\nonumber\\
&&T(\mu)=\frac{1}{(1-Z)^2}\sum_{n=0}^{D-2}T_n\left(\frac{1+Z}{1-Z}\right)^n,
\label{FT0}
\end{eqnarray}
where, as before, $F$ denotes generically $P,Q,R$ and $S$, and $F_n$ and $T_n$ are given by Eqs.~(\ref{PQRn}). We have thus obtained for the massless field  exactly the same result as in the limit $m\to 0$, except for the constant piece of $T(\mu)$ in Eq.~(\ref{T}). Remember that such a constant piece arises because of a mass term, which is lacking here. 

Due to the absence of the constant piece, the behavior at large distances changes dramatically with respect to the case of small masses: at $Z\ll -1$, we have
\begin{equation}
P,Q,R,S,T\sim Z^{-2}.
\label{lardis0}
\end{equation}
Therefore, {\emph{the stress tensor two-point function of the massless field vanishes at long distances. There is a ``discontinuity'' at $m=0$: taking the limit $m\to 0$ is not the same as setting $m=0$}. 

One might think that this discontinuity is due to our choice of states for both the massive and the massless fields, and that by choosing appropriate states the discontinuity would disappear. However, this does not seem to be the case. In Ref.~\cite{kirsten93}, the expectation value $_\alpha\langle 0|T_{ab}| 0\rangle_\alpha$ was computed for all values of $\alpha$. In the limit $\alpha\to 0$ the result is de Sitter-invariant and it disagrees with the limit $m\to 0$, similarly to what we have just seen for the stress tensor two-point function. Furthermore, at late times one has
\begin{equation*}
_\alpha\langle 0| T_{ab}| 0\rangle_\alpha\to \lim_{\alpha\to 0}{}_\alpha\langle 0| T_{ab}| 0\rangle_\alpha.
\end{equation*}
That is, $\lim_{\alpha\to 0}{}_\alpha\langle  0| T_{ab}| 0\rangle_\alpha$ is an attractor at late times. As is well-known, the same behavior occurs for the massive field in the Bunch-Davies vacuum. This means that, in the case of the expectation value of $T_{ab}$, the discontinuity at $m=0$ appears at late times regardless of the state of the quantum field. It is plausible that something similar happens for the stress tensor two-point function.

It is important to note that we are comparing the massless, minimally coupled field with a very particular limit in parameter space. In general, in $dS_D$ (and in any curved spacetime) the Klein-Gordon equation (\ref{kg}) includes the term $\xi R \phi$, where $R$ is the Ricci scalar and $\xi$ is a parameter called the coupling to the curvature. In this paper we have considered only minimally coupled fields, i.e., $\xi=0$, and we have compared the case $\xi=0,m=0$ with the case $\xi=0,m\to 0$. In Ref.~\cite{kirsten93}, the expectation value of the stress tensor in the case $\xi=0,m=0$ was compared with the same expectation value in the more general limit $\xi\to 0,m\to 0$. It was found that such a limit is not unique. In particular, it was shown that, if the limit is taken in such a way that $\xi R/m^2\to -2$, the discontinuity disappears. It is reasonable to think that the stress tensor two-point function will exhibit a similar behavior.

We close this subsection by mentioning that the stress tensor two-point function of a massless minimally coupled scalar field in $dS_4$ was also computed in Ref.~\cite{roura99b}. Although in qualitative agreement, the result of that reference differs from the results presented here because the covariant derivatives of the unit vectors were not properly taken into account there.


\subsection{Large distances}
\label{subseclar}

In the previous subsections we have studied the stress tensor fluctuations in the cases of small masses and strictly zero mass. In both cases we have paid attention to the long-distance ($Z\ll -1$) regime. Here we investigate the long-distance behavior of the stress tensor fluctuations of a massive field, without making assumptions on the smallness of $m$. 

Using the properties of the hypergeometric function \cite{abramowitz72}, one sees that the Wightman function (\ref{wig}) can be expanded as
\begin{equation}
G(\mu)=g_+(-Z)^{-h_+}+g_-(-Z)^{-h_-}+O(Z^{-\min (h_+,h_-)-1}),
\label{wigld}
\end{equation}}
where
\begin{equation}
g_\pm =\frac{2^{h_\pm}}{(4\pi)^{D/2}}\frac{\Gamma(h_\pm)\Gamma(h_\mp-h_\pm)}{\Gamma(\frac{D}{2}-h_\pm)},
\label{gpm}
\end{equation}
and where $h_{\pm}$ is given in (\ref{hc}). Now, there are two possible situations, depending on the size of $m$ compared to $(D-1)/2$. If $m<(D-1)/2$, the parameters $h_\pm$ are real, with $h_+>h_-$. Then, the first term in (\ref{wigld}) is subleading, and the equation simplifies to
\begin{equation}
G(\mu)=g_-(-Z)^{-h_-}+O(Z^{-\min (h_+,h_-+1)}).
\end{equation}
After substitution in (\ref{PQR}), one obtains the long-distance behavior of the stress tensor two-point function. The result is
\begin{equation}
P,Q,T\sim Z^{-2h_-}\qquad R\sim Z^{-2h_--1}\qquad S\sim Z^{-2h_--2}.
\label{lardis1}
\end{equation}
This is consistent with what we have seen in Sec.~\ref{subsecsma}: if we let $m\to 0$, it follows from (\ref{hc}) that $h_- \simeq m^2/(D-1)\to 0$, and the leading contribution to the stress tensor two-point function is a constant. In fact, the coefficients of the powers of $Z$ that have been omitted in the equation above all vanish when $m\to 0$ except those of $S$ and $T$. Therefore, the only function that tends to a constant is $T$, as has been seen in \ref{subsecsma}. Besides, $R$ decays more quickly than $Z^{-1}$, also consistently with that subsection. The parameter $h_-$ increases with the mass as long as $m<(D-1)/2$. Hence, the fluctuations decay faster with the distance as the mass increases.

In the opposite case, namely $m>(D-1)/2$, the parameters $h_\pm$ are complex, and so are $g_\pm$. According to Eqs.~(\ref{hc}) and (\ref{gpm}), we can write
\begin{equation*}
h_\pm=\frac{1}{2}(D-1)\pm i\alpha\qquad g_\pm=ge^{\pm i\delta},
\end{equation*}
where $\alpha,g$ and $\delta$ are positive real numbers. In terms of these parameters, Eq.~(\ref{wigld}) takes the form
\begin{equation}
G(\mu)=2g\cos[\alpha\ln(-Z)-\delta](-Z)^{-\frac{D-1}{2}}+O(Z^{-\frac{D+1}{2}}).
\end{equation}
Substituting this equation in (\ref{PQR}), we obtain the following result for the long-distance behavior of the stress tensor two-point function:
\begin{equation}
P,Q,T\sim Z^{-D+1}\qquad R\sim Z^{-D}\qquad S\sim Z^{-D-1},
\label{lardis2}
\end{equation}
where the coefficients that have been omitted are oscillating functions of $Z$. Hence, for $m>(D-1)/2$, the way in which the stress tensor fluctuations decay with the distance does not depend on the mass of the field. This is different from the situation in Minkowski spacetime, where the Wightman function decays with the distance like $e^{-m\mu}$, and accordingly the stress tensor fluctuations decay like $e^{-2m\mu}$. 

Naively, one might expect to recover the results of Minkowski spacetime when the Compton wavelength $1/m$ is much smaller than the Hubble radius (which here is set to one). Then one would find a contradiction with our results above for $m>(D-1)/2$. However, the correct expectation is that one should recover the results of Minkowski spacetime when both the Compton wavelength and the physical distance between the two points under consideration are smaller than the Hubble radius. This is not the case we are dealing with here, where we assume $Z\ll -1$, which means that the physical distance is much larger than the Hubble radius.

\section{Timelike separated points}
\label{sec:timelike}

In the previous two sections we have been discussing the properties of the stress tensor two-point function for spacelike separated points. Here we extend these results to timelike separated points.

The Wightman function of the free scalar field in the Bunch-Davies vacuum for timelike separated points, $Z>1$, is \cite{birrell94}
\begin{equation}
G^+=c_{m,D}F(h_+,h_-;\frac{D}{2};\frac{1+ \tilde Z}{2}),
\label{wigtime}
\end{equation}
where $F$ is the hypergeometric function, and $c_{m,D}$ and $h_\pm$ are given in Eq.~(\ref{hc}). Thus, $G^+$ has the same form as for spacelike separated points, except that now it depends on $\tilde Z$ instead of $Z$. The biscalar $\tilde Z$ is defined as
\begin{equation}
 \tilde Z(x,x')=Z(x,x')\mp i\epsilon,
\label{tilz}
\end{equation}
where $\epsilon$ is a positive real infinitesimal, and the minus (plus) sign is chosen when $x$ is to the future (past) of $x'$. This $\mp i\epsilon$ term is important because the hypergeometric function, $F(a,b;c;z)$, has a branch cut at $z>1$ along the real axis. 

Introducing the biscalar $\tilde\mu\equiv\cos^{-1}\tilde Z$, the Wightman function (\ref{wigtime}) can also be written as
\begin{equation}
G^+=G(\tilde \mu),
\label{wigtim2}
\end{equation}
where $G$ is the function defined in (\ref{wig}). From (\ref{muz2}) and the discussion below that equation it follows that
\begin{equation}
\tilde\mu(x,x')=\pm\mu(x,x')+\epsilon,
\end{equation}
where the plus (minus) sign is chosen when $x$ is to the future (past) of $x'$. Unlike in Eq.~(\ref{muz2}), here we are keeping the leading term that vanishes when $\epsilon\to 0^+$, because this limit has to be taken after evaluating the hypergeometric function at $\tilde Z$. Recall that the geodesic distance between timelike separated points is imaginary, as has been emphasized under Eq.~(\ref{geo}). 

Note that $\tilde\mu$ is not invariant under the full de Sitter group. For example, under the antipodal transformation it transforms as 
\begin{equation}
\tilde\mu(\alpha(x),\alpha(x'))=\tilde\mu^*(x,x').
\end{equation}
This is because, if $x$ is to the future of $x'$, then $\alpha(x)$ is to the past of $\alpha(x')$. Also, due to the branch cut of the hypergeometric function, $G(\tilde\mu)\ne G(\tilde\mu^*)$ and, hence, the Wightman function in the Bunch-Davies vacuum for timelike separated points is not invariant under the full de Sitter group. This is not surprising: for timelike separated points, $[\phi(x),\phi(x')]$ does not necessarily vanish. If this commutator does not vanish, the product $\phi(x)\phi(x')$ is not a Hermitian operator and, in consequence, the Wightman function does not have to be a de Sitter-invariant biscalar.

The stress tensor two-point function in the Bunch-Davies vacuum for timelike separated points is obtained by substitution of (\ref{wigtim2}) in (\ref{flu3}). Using (\ref{cov}), we find
\begin{eqnarray}
F_{abc'd'}&=&P(\tilde\mu)n_an_bn_{c'}n_{d'}+Q(\tilde\mu)(n_an_bg_{c'd'}+n_{c'}n_{d'}g_{ab})\nonumber\\
&&+R(\tilde\mu)(n_an_{c'}g_{bd'}+n_bn_{d'}g_{ac'}+n_an_{d'}g_{bc'}+n_bn_{c'}g_{ad'})\nonumber\\
&&+S(\tilde\mu)(g_{ac'}g_{bd'}+g_{bc'}g_{ad'})+T(\tilde\mu)g_{ab}g_{c'd'},
\label{flutim}
\end{eqnarray}
where $P,Q,R,S$ and $T$ are given by Eqs.~(\ref{PQR}) and (\ref{G1G2}). Therefore, the stress tensor two-point function for timelike separated points has the same form as for spacelike separated points, except that now the functions $P,Q,R,S$ and $T$ depend on $\tilde\mu$ instead of $\mu$. 

The explicit form of these functions in the particular cases of small masses and massless field is obtained simply by replacing $\mu$ by $\tilde\mu$ (or, equivalently, $Z$ by $\tilde Z$) in Eqs.~(\ref{F})-(\ref{PQRn}) and (\ref{FT0}). In fact, the functions of $Z$ appearing in those equations do not have any branch cut. Hence, the $\mp i\epsilon$ term in the definition of $\tilde Z$ is irrelevant, and those equations are valid themselves also for timelike separated points. This means that, for both $m=0$ and $m\to 0$, the stress tensor two-point function is also de Sitter-invariant for timelike separated points. The behavior at large separations, $Z\gg 1$, is obtained similarly, by replacing $Z$ by $\tilde Z$ in Eqs.~(\ref{lardis1}) and (\ref{lardis2}). There, strictly speaking we do have branch cuts (due to the non-integer exponents in (\ref{lardis1}) and the oscillating functions that have been omitted from (\ref{lardis2})), but the qualitative behavior is not affected by them.

In order to interpret physically these results, it is convenient to express them in terms if the geodesic distance $\mu$, because there is always a geodesic joining two timelike separated points. For $Z\gg 1$, we have $Z\sim \exp|\mu|$, as can be seen from (\ref{muz2}). Let us concentrate on the cases  $m\to 0$ and $m=0$. In the limit $m\to 0$, due to the constant piece in (\ref{T}), the fluctuations do not vanish at large separations. As we have already commented, this is different from what happens in Minkowski spacetime, where the fluctuations decay like $\mu^{-2D}$. In contrast, in the massless case it follows from (\ref{lardis0}) that the fluctuations decay exponentially with $|\mu|$, that is, much faster than in Minkowski spacetime.

\section{Discussion}
\label{sec:discussion}

In this paper we have studied the stress tensor two-point function of a quantum field in $dS_D$. We have given general constraints on this two-point function that hold for any quantum field theory in $dS_D$, assuming that the state of the field is de Sitter invariant and that the two points are spacelike separated. Next, we have derived explicit expressions for a free minimally coupled scalar field in the Bunch-Davies vacuum. These expressions take a simple form (in terms of elementary functions) in the particular cases of a small or vanishing mass as well as for large separations between the two points and general mass.
(In the case of a vanishing mass the Bunch-Davies vacuum does not exist, but one can consider a vacuum arbitrarily close to it, in the sense specified in the appendix.)
These simple expressions, which have been given in Sec.~\ref{sec:particular}, are the main result of this paper. The conclusions of that section can be summarized as follows. When the mass $m$ of the field is small, the correlations have a long range: they decay like an inverse power of the physical distance, with an exponent proportional to $m^2/H^2$. In particular, in the limit $m\to 0$ they do not vanish at long distances. In contrast, in the strictly massless case the stress tensor two-point function decays like the fourth power of the physical distance. There is thus a discontinuity in the massless limit.
Such a discontinuity may seem surprising at first sight, but it can be easily understood as follows. In the massive case, there is a contribution from the mass term in the stress tensor that does not vanish in the limit $m\to 0$ due to the infrared divergence of the Bunch-Davies vacuum.

By explicitly writing the dependence on the Hubble constant in our expressions for the stress-tensor two-point functions and taking the limit $H \to 0$ one can recover the flat-space limit and compare with the results of Ref.~\cite{martin00}, where they were computed in the Minkowski vacuum for a general mass and curvature coupling.
In fact, by generalizing Eqs.~(\ref{kg})-(\ref{G1G2}) one can straightforwardly (although somewhat tediously) extend our results to a general value of the  curvature coupling parameter. Furthermore, by an analytic continuation of the Hubble constant to $iH$ and the choice of a solution of Eq.~(\ref{kgw}) with appropriate boundary conditions in that case [rather than the solution given by Eq.~(\ref{wig})] one can also obtain the stress-tensor two-point functions for the natural vacua in anti-de Sitter. This will give the general result in terms of invariant bitensors for all maximally symmetric spacetimes (Minkowski, de Sitter and anti-de Sitter). By particularizing to the massless conformally coupled case, one can then compare with the results of Ref.~\cite{osborn00} for general conformal field theories. 

Next, we discuss how our results for the quantum correlations of the stress tensor can be used to obtain information on the quantum fluctuations of the gravitational field in de Sitter spacetime. Specifically, we will consider a perturbative quantization of the metric around a de Sitter background within the framework of an effective field theory approach to quantum gravity \cite{donoghue94b,burgess04}, and for simplicity we will restrict our attention to $D=4$ spacetime dimensions. Moreover, we will neglect the contributions from graviton loops in order to avoid the difficulties involved in constructing and dealing with diffeomorphism-invariant observables \cite{giddings06} that arise when those are taken into account. This can be naturally implemented in the context of a large $N$ expansion for $N$ matter fields (free scalar fields in our case) coupled to the gravitational field, where one introduces a rescaled coupling constant $\bar l_\mathrm{p}=l_\mathrm{p}\sqrt N$ (with $l_\mathrm{p}$ being the usual Planck length) and expands in powers of $1/N$ keeping $\bar l_\mathrm{p}$ fixed. Neglecting graviton loops amounts then to considering only the leading order in $1/N$, i.e.\ the first non-vanishing order that contributes to any quantity of interest. For instance, the leading contribution to the two-point function of the metric perturbations is of order $1/N$.

As a characterization of the quantum metric fluctuations at order $1/N$, one can study the two-point correlation functions of the linearized Riemann tensor including the effects of matter loops. It should be emphasized that we are not referring to transition matrix elements but to the correlation functions appearing in the closed time path (also known as \emph{in-in}), formalism \cite{schwinger61,chou85,jordan86,calzetta87,campos94}. They can be obtained either from a purely quantum field theoretical calculation \cite{weinberg05} or within the framework of \emph{stochastic gravity} \cite{calzetta94,martin99a,martin99b,hu03a,hu08}, which provides an equivalent way of obtaining  the result at order $1/N$ \cite{hu04b,tomboulis77}. The Riemann tensor of the background de Sitter spacetime with appropriately raised indices is given by $R_{ab}^{\ \ cd}= (R/6)\,\delta_{[a}^{\ [c}\delta_{b]}^{\ d]}$ and its linear perturbation is, therefore, gauge invariant since its Lie derivative with respect to an arbitrary vector vanishes \cite{stewart90}. In fact, both the Ricci tensor and the Weyl tensor (which vanishes in the background spacetime) are separately gauge invariant at the linear level, and together they entirely determine the Riemann tensor at that order. This is actually enough to guarantee that the connected part of their two-point correlation functions at order $1/N$ is gauge invariant.
One could alternatively consider the two-point function of other gauge-invariant quantities for the linearized metric perturbations, such as those commonly employed in the study of cosmological perturbations. Instead, here we consider the Riemann tensor because besides being a gauge-invariant object with a simple geometric meaning and completely characterizing the local geometry, the de Sitter invariance of its correlation functions can be directly analyzed, especially when using a manifestly invariant formalism like the one employed in this paper.
Furthermore, from our results for the stress-tensor two-point functions one can immediately obtain the two-point functions of the Ricci tensor to leading order in the Planck length (and order $1/N$), as we will see below.

The lowest-order non-vanishing contribution to the two-point function of the Ricci tensor, or equivalently the Einstein tensor, is quartic in $\bar l_\mathrm{p}$ and (at order $1/N$) comes from the one-loop contributions of the matter fields. Both from the stochastic gravity formalism \cite{hu04b,hu08} or from the Einstein equation to linear order in the metric perturbations as an equation for operators in the Heisenberg picture (including the matter fields as well as the metric perturbations), one obtains the following result for the two-point function of the Einstein tensor at that order:
\begin{equation}
\left\langle G^{a\,(1)}_{\ b}(x)\,G^{c'\,(1)}_{\ d'}(x') \right\rangle
-\left\langle G^{a\,(1)}_{\ b}(x)\right\rangle \left\langle G^{c'\,(1)}_{\ d'}(x')\right\rangle
= (8\pi)^2\, \frac{\bar l_\mathrm{p}^{\,4}}{N}\, F^{a\ c'}_{\ b\ d'}(x,x'),
\label{eq:correl1}
\end{equation}
where $G^{a(1)}_{\ b}$ is the linearized Einstein tensor, and $F^{a\ c'}_{\ b\ d'}$ is the stress tensor two-point function that has been computed in this paper, with appropriately raised indices. Therefore, the stress tensor two-point function in $dS_4$ gives the leading order contribution to the two-point function of the linearized Einstein tensor in an expansion in powers of $1/N$ and $\bar l_\mathrm{p}$. Note that in general this two-point function would receive additional contributions of the same order from the finite part of the local counterterms quadratic in the curvature which are required in order to renormalize the divergences of the one-loop contribution, but at this order these contributions turn out to vanish for the particular case of a de Sitter background.

On the other hand, even though the two-point function of the Weyl tensor also gets contributions of order $\bar l_\mathrm{p}^{\,4}$ and higher from matter loops, it gets in addition a non-vanishing tree-level contribution of order $\bar l_\mathrm{p}^{\,2}$ even in the absence of matter loops, which was calculated in Ref.~\cite{kouris01}. In contrast with the correlation functions of the Ricci tensor, the contributions of order $\bar l_\mathrm{p}^{\,4}$ to the two-point function of the Weyl tensor cannot be immediately obtained from the stress-tensor two-point function. Nevertheless, when combined with the tree-level result for the correlations of the Weyl tensor in Ref.~\cite{kouris01}, our current result for the correlations of the Ricci tensor completely characterizes the quantum fluctuations of the full Riemann tensor at order $1/N$ and leading order in the Planck length. It should be pointed out that the two-point function corresponding to the cross-correlation of the Ricci and Weyl tensors is of order $\bar l_\mathrm{p}^{\,4}$ (rather than $\bar l_\mathrm{p}^{\,3}$). This can be interpreted as the absence of correlations between the tree-level fluctuations of the Weyl tensor, with an amplitude of order $\bar l_\mathrm{p}$,
and the fluctuations of the Ricci tensor, with an amplitude of order $\bar l_\mathrm{p}^{\,2}$.

Note that in contrast to the existing calculations of loop corrections to the spectrum of cosmological perturbations in inflationary models, where the self-interaction of the matter fields often plays an important role \cite{sloth06,sloth07,seery07}, the only interaction vertices associated with the free fields that we have considered here are those corresponding to their gravitational interaction. Moreover, we do not consider a non-vanishing homogeneous background configuration for the scalar fields (which would give also a tree-level contribution
to the stress-tensor two-point function \cite{roura08}) and the usual treatment for the correlations of the scalar-type metric perturbations \cite{maldacena03b,weinberg05,seery08,adshead09b} cannot be directly applied. On the other hand, the calculation for tensor metric perturbations should be equivalent (since the slow-roll geometry is usually approximated by de Sitter space in that case) but the existing results are either approximate \cite{urakawa08b} or restricted to the massless case \cite{adshead09b}. Furthermore, considering the correlations of the Riemann tensor and expressing them in terms of maximally invariant tensors is very useful in order to analyze whether de Sitter invariance is disrupted by loop corrections.

We close this section with a few remarks concerning possible lines of future research in connection with our work. An aspect that deserves further study is analyzing how generic our main results for light fields (namely, the existence of long-range stress-tensor correlations and the discontinuity of the massless limit) are when considering general states. For instance, for a wide range of masses and curvature-coupling parameters it has been shown that given a generic well-behaved and spatially isotropic initial state, the expectation value of the stress-tensor operator on a de Sitter background tends at late times to the same value as the expectation value for the Bunch-Davies vacuum \cite{anderson00}. Moreover, for spatially isotropic perturbations this conclusion remains true even when the back-reaction on the background geometry is taken into account \cite{isaacson91,rogers92,busch92,perez-nadal08a,perez-nadal08b}. This can be qualitatively understood as a consequence of the exponential redshift rendering negligible, after a sufficiently long time, the contribution to the stress tensor of any initial excitation of the Bunch-Davies vacuum. Hence, we expect that a similar conclusion will apply to the stress-tensor correlations and its main features,
but it would be interesting to check it explicitly.
Secondly, in future work we plan to compute the one-loop contribution (of order $\bar l_\mathrm{p}^4$) to the two-point function of the Weyl tensor, which unlike for the Ricci tensor cannot be immediately obtained from the stress tensor two-point function, and investigate the de Sitter invariance of the one-loop contributions to the correlation function of the full Riemann tensor and its implications.

\section*{Acknowledgments}

It is a pleasure to thank Esteban Calzetta for valuable discussions while one of us was visiting the University of Buenos Aires, some of which motivated us to pursue this project. We are also grateful to Jaume Garriga for interesting conversations and the participants of the 14th Peyresq Cosmology Meeting for useful comments. This work has been partly supported by the Research Projects MEC FPA2007-66665C02-02, CPAN CSD2007-00042, within the program Consolider-Ingenio 2010, and AGAUR 2009SGR00168. In addition, during the initial stages of this project A.~R.\ was partly supported by a DOE grant. G.~P.\ acknowledges support through the FPI grant BES-2005-10146.

\appendix

\section{Vacuum states}
\label{app:vacua}

In this appendix we summarize the quantization of a free, minimally coupled scalar field in $dS_D$. We pay special attention to the massless case, where we reexamine the issue of the non-existence of a de Sitter-invariant vacuum, previously addressed in Ref.~\cite{allen85, kirsten93}, and prove Eq.~(\ref{O4Bun1}). For simplicity, we will work in $D=4$ dimensions of spacetime.

A coordinate system $\{\eta,\Omega\}$ can be chosen for $dS_4$ such that the line element takes the form
\begin{equation}
ds^2=\sin^{-2}\eta[-d\eta^2+d\Omega^2].
\end{equation}
where $\eta\in (0,\pi)$ is the so-called conformal time, $\Omega$ is a set of angles of $S^3$, and $d\Omega^2$ is the line element of $S^3$. In these coordinates, the scalar field can be expanded in spherical harmonics on $S^3$, 
\begin{equation}
\phi(\eta,\Omega)=\sum_{L,M}[a_{LM}U_{LM}(\eta)Y_{LM}(\Omega)+a_{LM}^\dagger U_{LM}^*(\eta)Y_{LM}^*(\Omega)],
\end{equation}
where the spherical harmonics $Y_{LM}(\Omega)$ are normalized eigenfunctions of the Laplacian on $S^3$, $L=0,\dots,\infty$, and $M$ represents two indices, $M_1$ and $M_2$, with $M_1=-L,\dots,L$ and $M_2=-M_1,\dots,M_1$. In order for the scalar field to satisfy the Klein-Gordon equation (\ref{kg}), the modes $U_{LM}$ have to be particular solutions of the equation 
\begin{equation}
\ddot{U}_{LM}(\eta)-2\cot\eta\,\dot{U}_{LM}(\eta) +\omega_{L}^2(\eta)U_{LM}(\eta)=0,
\label{modeq}
\end{equation}
where $\omega_{L}^2(\eta)=L(L+2)+m^2\sin^{-2}\eta$. If these particular solutions satisfy the Wronskian condition
\begin{equation}
U_{LM}(\eta)\dot U_{LM}^*(\eta)-U_{LM}^*(\eta)\dot U_{LM}(\eta)=i\sin^2\eta,
\label{wro}
\end{equation}
then the canonical commutation relations between the field and its conjugate momentum imply that $a_{LM}^\dagger$ and $a_{LM}$ are creation and annihilation operators, 
\begin{equation}
[a_{LM},a_{L'M'}^\dagger]=\delta_{LL'}\delta_{MM'}.
\end{equation}
The vacuum $|0\rangle$ is then defined as the state annihilated by all the annihilation operators,
\begin{equation}
a_{LM}|0\rangle=0 \qquad \forall L,M.
\end{equation}
In spite of the Wronskian condition (\ref{wro}), there is freedom in choosing the modes $U_{LM}$. Different choices give rise to different vacua. The Bunch-Davies vacuum, $|0\rangle_{BD}$, is associated with the following choice \cite{chernikov68,birrell94}
\begin{equation}
U_{LM}^{BD}(\eta)=A_L\sin^{3/2}\eta[P^{\lambda}_\nu(-\cos\eta)-\frac{2i}{\pi}Q^{\lambda}_\nu(-\cos\eta)],
\label{BD}
\end{equation}
where $P^{\lambda}_\nu$ and $Q^{\lambda}_\nu$ are Legendre functions on the cut, $\lambda=(9/4-m^2)^{1/2}$, $\nu=L+1/2$, and the normalization constants are 
\begin{equation}
A_L=\frac{\sqrt{\pi}}{2}e^{i\lambda\pi/2}\left[\frac{\Gamma(L-\lambda+3/2)}{\Gamma(L+\lambda+3/2)}\right]^{1/2}.
\label{AL}
\end{equation}
The Bunch-Davies vacuum is de Sitter-invariant. Note that this state is only defined for massive fields, $m>0$, because $A_0$ does not exist for $m=0$. In fact, in the massless case there is no de Sitter-invariant vacuum, as was shown in Ref.~\cite{allen85}. One can instead consider the vacua $|0\rangle_\alpha$, associated with the following modes \cite{allen87b}:
\begin{eqnarray}
&&U_{LM}^{\alpha}(\eta)=\lim_{m\to 0}U_{LM}^{BD}(\eta)\qquad {\text{for }} L>0\nonumber\\
&&U_{0}^{\alpha}(\eta)= -i\alpha(\eta-\frac{1}{2}\sin 2\eta-\frac{\pi}{2})+\frac{1}{4\alpha},
\label{O4}
\end{eqnarray}
where $\alpha\in (0,\infty)$. These vacua are invariant under the $O(4)$ subgroup of $O(4,1)$, i.e., the subgroup of isometries that leave invariant the hypersurfaces of constant $\eta$. They are also invariant under time reversal. (These vacua should not be confused with the so-called \emph{alpha vacua}, which are defined for massive fields and whose Wightman function diverges for antipodal points.)

In what follows we concentrate on the massless case. As was pointed out in Ref.~\cite{kirsten93}, the non-existence of the Bunch-Davies vacuum when $m=0$ is better understood by looking at the example of the harmonic oscillator in Quantum Mechanics. The general solution of the equation of motion of a harmonic oscillator of frequency $\omega$ can be written in the form 
\begin{equation}
q(t)=U(t)a+ U^*(t)a^\dagger,
\label{har}
\end{equation}
where $a^\dagger$ and $a$ are creation and annihilation operators, and $U(t)$ is a particular solution of the equation of motion. If it is chosen as
\begin{equation}
U(t)=\frac{1}{\sqrt{2\omega}}{e^{-i\omega t}},
\label{u}
\end{equation}
then the ground state $|\psi\rangle$ of the harmonic oscillator satisfies $a|\psi\rangle=0$. Now, $U(t)$ does not exist when $\omega=0$. This is equivalent to what happens to $U_{0}^{BD}$ when $m=0$. But in this case the interpretation is clear: a harmonic oscillator of zero frequency is a free particle. And the ground state of the free particle, which should be annihilated by the momentum operator, $p|\psi\rangle=0$, does not exist, because a constant wave function cannot be normalized. Similarly \cite{kirsten93}, the wave functional of the Bunch-Davies vacuum becomes non-normalizable in the limit $m\to 0$.

Even though the ground state of the free particle does not exist in Quantum Mechanics, there are states which are ``arbitrarily close'' to it. An example is a Gaussian wave function with very large width. Similarly, one can expect to find states for the massless field in $dS_4$ which are ``arbitrarily close'' to being de Sitter-invariant. This is indeed true, in the following (restricted) sense: even though the vacua $|0\rangle_\alpha$ are not de Sitter-invariant, there is a certain class of observables whose expectation values are de Sitter-invariant in the limit $\alpha\to 0$. 

Let us see this. For any mass $m$, and in any vacuum state $|0\rangle$, the Wightman function of the free scalar field in $dS_4$ is expanded in terms of the modes as
\begin{equation}
G^+(\eta,\Omega;\eta',\Omega')=\sum_{L,M}Y_{LM}(\Omega)Y_{LM}^*(\Omega')U_{LM}(\eta)U_{LM}^*(\eta').
\label{wigmod}
\end{equation}
If we want to compute the Wightman function, $G^{+}_\alpha$, of a massless field in the state $|0\rangle_\alpha$, we have to replace $U_{LM}$ by $U_{LM}^\alpha$ in this equation. If, instead, we want to compute the Wightman function, $G^{+}_{BD}$, of a massive field in the Bunch-Davies vacuum, we will use $U_{LM}^{BD}$. Let us consider the limit $\alpha\to 0$ for a massless field in the state $|0\rangle_\alpha$. It follows from (\ref{O4}) that the zero mode $U_{0}^\alpha$ diverges in this limit, and so does the Wightman function $G^{+}_\alpha$. However, the derivative of the zero mode, $\dot U_{0}^\alpha$, vanishes. Therefore, $\partial_\eta\partial_{\eta'}G^{+}_\alpha$ is finite in the limit $\alpha\to 0$ and, besides, the term with $L=0$ does not contribute to it. On the other hand, consider the limit $m\to 0$ for a massive field in the Bunch-Davies vacuum. It follows from (\ref{AL}) that the zero mode $U_{0}^{BD}$ diverges. This is the origin of the infrared divergence of $G^{+}_{BD}$ found in Sec.~\ref{subsecsma}. However, the derivative of the zero mode, $\dot U_{0}^{BD}$, vanishes. Hence, $\partial_\eta\partial_{\eta'}G^{+}_{BD}$ is finite in the limit $m\to 0$, and the term with $L=0$ does not contribute to it. Using Eqs.~(\ref{O4}) and (\ref{wigmod}) we find
\begin{equation}
\lim_{\alpha\to 0}\partial_\eta\partial_{\eta'}G^{+}_\alpha=\lim_{m\to 0}\partial_\eta\partial_{\eta'}G^{+}_{BD}.
\end{equation}
Finally, the spherical harmonic $Y_{00}$ is independent of $\Omega$. Therefore, whenever we take derivatives of the Wightman function (\ref{wigmod}) with respect to $\Omega$ or $\Omega'$ the term with $L=0$ will not contribute, and a result similar to the above equation will apply. From all this we conclude that
\begin{equation}
\lim_{\alpha\to 0}\nabla_a\nabla_{b'}G^{+}_\alpha=\lim_{m\to 0}\nabla_a\nabla_{b'}G^{+}_{BD}.
\label{O4Bun}
\end{equation}
Since the right-hand side is de Sitter-invariant, this equation implies that $\nabla_a\nabla_{b'}G^{+}_\alpha$ is de Sitter-invariant in the limit $\alpha\to 0$. The expectation value in the vacuum $|0\rangle_\alpha$ of any observable that only depends on derivatives of the field can be expressed in terms of $\nabla_a\nabla_{b'}G^{+}_\alpha$, because $|0\rangle_\alpha$ is a Gaussian state. In consequence, such expectation value will be de Sitter-invariant in the limit $\alpha\to 0$. Hence, we conclude that, even though the states $|0\rangle_\alpha$ are not de Sitter-invariant, the expectation values of certain observables can be taken arbitrarily close to a de Sitter-invariant value by choosing $\alpha$ sufficiently small.

Note that the conclusions above apply only to a certain class of observables, not all of them. For example, $\nabla_a G^{+}_\alpha$ does not satisfy an equation like (\ref{O4Bun}). The reason is that, when computing this bitensor using Eq.~(\ref{wigmod}), the contribution from the term with $L=0$ is finite, but it does not vanish. Hence, one cannot use the argument that led to Eq.~(\ref{O4Bun}). It is also worth noting that $|0\rangle_\alpha$ is not the only family of states satisfying (\ref{O4Bun}). For instance, consider the family of coherent states $|z\rangle_\alpha$, defined by $a_{LM}|z\rangle_\alpha=0$ for $L>0$ and $a_0|z\rangle_\alpha=z|z\rangle_\alpha$, where $z$ is a complex number, and the annihilation operators are those associated with the $O(4)$ invariant vacua $|0\rangle_\alpha$. It is not hard to see that this family of states also satisfies Eq.~(\ref{O4Bun}) for all values of $z$.



\begin{thebibliography}{69}
\expandafter\ifx\csname natexlab\endcsname\relax\def\natexlab#1{#1}\fi
\expandafter\ifx\csname bibnamefont\endcsname\relax
  \def\bibnamefont#1{#1}\fi
\expandafter\ifx\csname bibfnamefont\endcsname\relax
  \def\bibfnamefont#1{#1}\fi
\expandafter\ifx\csname citenamefont\endcsname\relax
  \def\citenamefont#1{#1}\fi
\expandafter\ifx\csname url\endcsname\relax
  \def\url#1{\texttt{#1}}\fi
\expandafter\ifx\csname urlprefix\endcsname\relax\def\urlprefix{URL }\fi
\providecommand{\bibinfo}[2]{#2}
\providecommand{\eprint}[2][]{\url{#2}}

\bibitem{mukhanov05}
\bibinfo{author}{\bibfnamefont{V.~F.} \bibnamefont{Mukhanov}},
Ê\emph{\bibinfo{title}{Physical Foundations of Cosmology}}
Ê(\bibinfo{publisher}{Cambridge University Press},
Ê\bibinfo{address}{Cambridge}, \bibinfo{year}{2005}).

\bibitem{weinberg08}
\bibinfo{author}{\bibfnamefont{S.}~\bibnamefont{Weinberg}},
Ê\emph{\bibinfo{title}{Cosmology}} (\bibinfo{publisher}{Oxford University
ÊPress}, \bibinfo{address}{New York}, \bibinfo{year}{2008}).

\bibitem{mukhanov92}
\bibinfo{author}{\bibfnamefont{V.~F.} \bibnamefont{Mukhanov}},
Ê\bibinfo{author}{\bibfnamefont{H.~A.} \bibnamefont{Feldman}},
Ê\bibnamefont{and} \bibinfo{author}{\bibfnamefont{R.~H.}
Ê\bibnamefont{Brandenberger}}, \bibinfo{journal}{Phys. Rep.}
Ê\textbf{\bibinfo{volume}{215}}, \bibinfo{pages}{203} (\bibinfo{year}{1992}).

\bibitem{mukhanov97}
\bibinfo{author}{\bibfnamefont{V.~F.} \bibnamefont{Mukhanov}},
Ê\bibinfo{author}{\bibfnamefont{L.~R.~W.} \bibnamefont{Abramo}},
Ê\bibnamefont{and} \bibinfo{author}{\bibfnamefont{R.~H.}
Ê\bibnamefont{Brandenberger}}, \bibinfo{journal}{Phys. Rev. Lett.}
Ê\textbf{\bibinfo{volume}{78}}, \bibinfo{pages}{1624} (\bibinfo{year}{1997}).

\bibitem{abramo97}
\bibinfo{author}{\bibfnamefont{L.~R.~W.} \bibnamefont{Abramo}},
Ê\bibinfo{author}{\bibfnamefont{R.~H.} \bibnamefont{Brandenberger}},
Ê\bibnamefont{and} \bibinfo{author}{\bibfnamefont{V.~F.}
Ê\bibnamefont{Mukhanov}}, \bibinfo{journal}{Phys. Rev. D}
Ê\textbf{\bibinfo{volume}{56}}, \bibinfo{pages}{3248} (\bibinfo{year}{1997}).

\bibitem{abramo99}
\bibinfo{author}{\bibfnamefont{L.~R.~W.} \bibnamefont{Abramo}}
Ê\bibnamefont{and} \bibinfo{author}{\bibfnamefont{R.~P.}
Ê\bibnamefont{Woodard}}, \bibinfo{journal}{Phys. Rev. D}
Ê\textbf{\bibinfo{volume}{60}}, \bibinfo{pages}{044010}
Ê(\bibinfo{year}{1999}).

\bibitem{losic05}
\bibinfo{author}{\bibfnamefont{B.}~\bibnamefont{Losic}} \bibnamefont{and}
Ê\bibinfo{author}{\bibfnamefont{W.~G.} \bibnamefont{Unruh}},
Ê\bibinfo{journal}{Phys. Rev. D} \textbf{\bibinfo{volume}{72}},
Ê\bibinfo{pages}{123510} (\bibinfo{year}{2005}).

\bibitem{losic08}
\bibinfo{author}{\bibfnamefont{B.}~\bibnamefont{Losic}} \bibnamefont{and}
Ê\bibinfo{author}{\bibfnamefont{W.~G.} \bibnamefont{Unruh}},
Ê\bibinfo{journal}{Phys. Rev. Lett.} \textbf{\bibinfo{volume}{101}},
Ê\bibinfo{pages}{111101} (\bibinfo{year}{2008}).

\bibitem{tsamis96a}
\bibinfo{author}{\bibfnamefont{N.~C.} \bibnamefont{Tsamis}} \bibnamefont{and}
Ê\bibinfo{author}{\bibfnamefont{R.~P.} \bibnamefont{Woodard}},
Ê\bibinfo{journal}{Nucl. Phys. B} \textbf{\bibinfo{volume}{474}},
Ê\bibinfo{pages}{235} (\bibinfo{year}{1996}).

\bibitem{tsamis97}
\bibinfo{author}{\bibfnamefont{N.~C.} \bibnamefont{Tsamis}} \bibnamefont{and}
Ê\bibinfo{author}{\bibfnamefont{R.~P.} \bibnamefont{Woodard}},
Ê\bibinfo{journal}{Ann. Phys. (NY)} \textbf{\bibinfo{volume}{253}},
Ê\bibinfo{pages}{1} (\bibinfo{year}{1997}).

\bibitem{unruh98}
\bibinfo{author}{\bibfnamefont{W.~G.} \bibnamefont{Unruh}}
Ê(\bibinfo{year}{1998}), \eprint{arXiv:gr-qc/9802323}.

\bibitem{abramo02a}
\bibinfo{author}{\bibfnamefont{L.~R.~W.} \bibnamefont{Abramo}}
Ê\bibnamefont{and} \bibinfo{author}{\bibfnamefont{R.~P.}
Ê\bibnamefont{Woodard}}, \bibinfo{journal}{Phys. Rev. D}
Ê\textbf{\bibinfo{volume}{65}}, \bibinfo{pages}{043507}
Ê(\bibinfo{year}{2002}).

\bibitem{geshnizjani02}
\bibinfo{author}{\bibfnamefont{G.}~\bibnamefont{Geshnizjani}} \bibnamefont{and}
Ê\bibinfo{author}{\bibfnamefont{R.}~\bibnamefont{Brandenberger}},
Ê\bibinfo{journal}{Phys. Rev. D} \textbf{\bibinfo{volume}{66}},
Ê\bibinfo{pages}{123507} (\bibinfo{year}{2002}).

\bibitem{geshnizjani05}
\bibinfo{author}{\bibfnamefont{G.}~\bibnamefont{Geshnizjani}} \bibnamefont{and}
Ê\bibinfo{author}{\bibfnamefont{R.}~\bibnamefont{Brandenberger}},
Ê\bibinfo{journal}{JCAP} \textbf{\bibinfo{volume}{04}},
Ê\bibinfo{pages}{006} (\bibinfo{year}{2005}).

\bibitem{tsamis05}
\bibinfo{author}{\bibfnamefont{N.~C.} \bibnamefont{Tsamis}} \bibnamefont{and}
Ê\bibinfo{author}{\bibfnamefont{R.~P.} \bibnamefont{Woodard}},
Ê\bibinfo{journal}{Class. Quant. Grav.} \textbf{\bibinfo{volume}{22}},
Ê\bibinfo{pages}{4171} (\bibinfo{year}{2005}).

\bibitem{garriga08}
\bibinfo{author}{\bibfnamefont{J.}~\bibnamefont{Garriga}} \bibnamefont{and}
Ê\bibinfo{author}{\bibfnamefont{T.}~\bibnamefont{Tanaka}},
Ê\bibinfo{journal}{Phys. Rev. D} \textbf{\bibinfo{volume}{77}},
Ê\bibinfo{pages}{024021} (\bibinfo{year}{2008}).

\bibitem{allen87a}
\bibinfo{author}{\bibfnamefont{B.}~\bibnamefont{Allen}} \bibnamefont{and}
Ê\bibinfo{author}{\bibfnamefont{M.}~\bibnamefont{Turyn}},
Ê\bibinfo{journal}{Nucl. Phys. B} \textbf{\bibinfo{volume}{292}},
Ê\bibinfo{pages}{813} (\bibinfo{year}{1987}).

\bibitem{antoniadis91}
\bibinfo{author}{\bibfnamefont{I.}~\bibnamefont{Antoniadis}} \bibnamefont{and}
Ê\bibinfo{author}{\bibfnamefont{E.}~\bibnamefont{Mottola}},
Ê\bibinfo{journal}{J. Math. Phys.} \textbf{\bibinfo{volume}{32}},
Ê\bibinfo{pages}{1037} (\bibinfo{year}{1991}).

\bibitem{hawking00b}
\bibinfo{author}{\bibfnamefont{S.~W.} \bibnamefont{Hawking}},
Ê\bibinfo{author}{\bibfnamefont{T.}~\bibnamefont{Hertog}}, \bibnamefont{and}
Ê\bibinfo{author}{\bibfnamefont{N.}~\bibnamefont{Turok}},
Ê\bibinfo{journal}{Phys. Rev. D} \textbf{\bibinfo{volume}{62}},
Ê\bibinfo{pages}{063502} (\bibinfo{year}{2000}).

\bibitem{higuchi03}
\bibinfo{author}{\bibfnamefont{A.}~\bibnamefont{Higuchi}} \bibnamefont{and}
Ê\bibinfo{author}{\bibfnamefont{R.~H.} \bibnamefont{Weeks}},
Ê\bibinfo{journal}{Class. Quant. Grav.} \textbf{\bibinfo{volume}{20}},
Ê\bibinfo{pages}{3005} (\bibinfo{year}{2003}).

\bibitem{kouris01}
\bibinfo{author}{\bibfnamefont{S.~S.} \bibnamefont{Kouris}},
Ê\bibinfo{journal}{Class. Quant. Grav.} \textbf{\bibinfo{volume}{18}},
Ê\bibinfo{pages}{4961} (\bibinfo{year}{2001}).

\bibitem{allen86}
\bibinfo{author}{\bibfnamefont{B.}~\bibnamefont{Allen}} \bibnamefont{and}
Ê\bibinfo{author}{\bibfnamefont{T.}~\bibnamefont{Jacobson}},
Ê\bibinfo{journal}{Comm. Math. Phys.} \textbf{\bibinfo{volume}{103}},
Ê\bibinfo{pages}{669} (\bibinfo{year}{1986}).

\bibitem{polyakov08}
\bibinfo{author}{\bibfnamefont{A.~M.} \bibnamefont{Polyakov}},
Ê\bibinfo{journal}{Nucl. Phys. B} \textbf{\bibinfo{volume}{797}},
Ê\bibinfo{pages}{199} (\bibinfo{year}{2008}).

\bibitem{alvarez09}
\bibinfo{author}{\bibfnamefont{E.}~\bibnamefont{Alvarez}} \bibnamefont{and}
Ê\bibinfo{author}{\bibfnamefont{R.}~\bibnamefont{Vidal}}
Ê(\bibinfo{year}{2009}), \eprint{arXiv:0907.2375 [hep-th]}.

\bibitem{akhmedov09}
\bibinfo{author}{\bibfnamefont{E.~T.} \bibnamefont{Akhmedov}}
Ê(\bibinfo{year}{2009}), \eprint{arXiv:0909.3722v3 [hep-th]}.

\bibitem{bunch78a}
\bibinfo{author}{\bibfnamefont{T.~S.} \bibnamefont{Bunch}} \bibnamefont{and}
Ê\bibinfo{author}{\bibfnamefont{P.~C.~W.} \bibnamefont{Davies}},
Ê\bibinfo{journal}{Proc. R. Soc. London A} \textbf{\bibinfo{volume}{360}},
Ê\bibinfo{pages}{117} (\bibinfo{year}{1978}).

\bibitem{birrell94}
\bibinfo{author}{\bibfnamefont{N.~D.} \bibnamefont{Birrell}} \bibnamefont{and}
Ê\bibinfo{author}{\bibfnamefont{P.~C.~W.} \bibnamefont{Davies}},
Ê\emph{\bibinfo{title}{Quantum fields in curved space}}
Ê(\bibinfo{publisher}{Cambridge University Press},
Ê\bibinfo{address}{Cambridge}, \bibinfo{year}{1994}).

\bibitem{allen85}
\bibinfo{author}{\bibfnamefont{B.}~\bibnamefont{Allen}},
Ê\bibinfo{journal}{Phys. Rev. D} \textbf{\bibinfo{volume}{32}},
Ê\bibinfo{pages}{3136} (\bibinfo{year}{1985}).

\bibitem{misner73}
\bibinfo{author}{\bibfnamefont{C.~W.} \bibnamefont{Misner}},
Ê\bibinfo{author}{\bibfnamefont{K.~S.} \bibnamefont{Thorne}},
Ê\bibnamefont{and} \bibinfo{author}{\bibfnamefont{J.~A.}
Ê\bibnamefont{Wheeler}}, \emph{\bibinfo{title}{Gravitation}}
Ê(\bibinfo{publisher}{Freeman}, \bibinfo{address}{San Francisco},
Ê\bibinfo{year}{1973}).

\bibitem{wald84}
\bibinfo{author}{\bibfnamefont{R.~M.} \bibnamefont{Wald}},
Ê\emph{\bibinfo{title}{General Relativity}} (\bibinfo{publisher}{The
ÊUniversity of Chicago Press}, \bibinfo{address}{Chicago},
Ê\bibinfo{year}{1984}).

\bibitem{synge60}
\bibinfo{author}{\bibfnamefont{J.~L.} \bibnamefont{Synge}},
Ê\emph{\bibinfo{title}{Relativity: the general theory}}
Ê(\bibinfo{publisher}{North-Holland}, \bibinfo{address}{Amsterdam},
Ê\bibinfo{year}{1960}).

\bibitem{spradlin01}
\bibinfo{author}{\bibfnamefont{M.}~\bibnamefont{Spradlin}},
Ê\bibinfo{author}{\bibfnamefont{A.}~\bibnamefont{Strominger}},
Ê\bibnamefont{and} \bibinfo{author}{\bibfnamefont{A.}~\bibnamefont{Volovich}}
Ê(\bibinfo{year}{2001}), \eprint{arXiv:hep-th/0110007}.

\bibitem{abramowitz72}
\bibinfo{author}{\bibfnamefont{M.}~\bibnamefont{Abramowitz}} \bibnamefont{and}
Ê\bibinfo{author}{\bibfnamefont{I.~A.} \bibnamefont{Stegun}},
Ê\emph{\bibinfo{title}{Handbook of mathematical functions}}
Ê(\bibinfo{publisher}{Dover}, \bibinfo{address}{New York},
Ê\bibinfo{year}{1972}).

\bibitem{allen87b}
\bibinfo{author}{\bibfnamefont{B.}~\bibnamefont{Allen}} \bibnamefont{and}
Ê\bibinfo{author}{\bibfnamefont{A.}~\bibnamefont{Folacci}},
Ê\bibinfo{journal}{Phys. Rev. D} \textbf{\bibinfo{volume}{35}},
Ê\bibinfo{pages}{3771} (\bibinfo{year}{1987}).

\bibitem{kirsten93}
\bibinfo{author}{\bibfnamefont{K.}~\bibnamefont{Kirsten}} \bibnamefont{and}
Ê\bibinfo{author}{\bibfnamefont{J.}~\bibnamefont{Garriga}},
Ê\bibinfo{journal}{Phys. Rev. D} \textbf{\bibinfo{volume}{48}},
Ê\bibinfo{pages}{567} (\bibinfo{year}{1993}).

\bibitem{roura99b}
\bibinfo{author}{\bibfnamefont{A.}~\bibnamefont{Roura}} \bibnamefont{and}
Ê\bibinfo{author}{\bibfnamefont{E.}~\bibnamefont{Verdaguer}},
Ê\bibinfo{journal}{Int. J. Theor. Phys.} \textbf{\bibinfo{volume}{38}},
Ê\bibinfo{pages}{3123} (\bibinfo{year}{1999}).

\bibitem{martin00}
\bibinfo{author}{\bibfnamefont{R.}~\bibnamefont{Mart\'{\i }n}}
Ê\bibnamefont{and}
Ê\bibinfo{author}{\bibfnamefont{E.}~\bibnamefont{Verdaguer}},
Ê\bibinfo{journal}{Phys. Rev. D} \textbf{\bibinfo{volume}{61}},
Ê\bibinfo{pages}{124024} (\bibinfo{year}{2000}).

\bibitem{osborn00}
\bibinfo{author}{\bibfnamefont{H.}~\bibnamefont{Osborn}} \bibnamefont{and}
Ê\bibinfo{author}{\bibfnamefont{G.~M.} \bibnamefont{Shore}},
Ê\bibinfo{journal}{Nucl. Phys. B} \textbf{\bibinfo{volume}{571}},
Ê\bibinfo{pages}{287} (\bibinfo{year}{2000}).

\bibitem{donoghue94b}
\bibinfo{author}{\bibfnamefont{J.~F.} \bibnamefont{Donoghue}},
Ê\bibinfo{journal}{Phys. Rev. D} \textbf{\bibinfo{volume}{50}},
Ê\bibinfo{pages}{3874} (\bibinfo{year}{1994}).

\bibitem{burgess04}
\bibinfo{author}{\bibfnamefont{C.~P.} \bibnamefont{Burgess}},
Ê\bibinfo{journal}{Living Rev. Rel.} \textbf{\bibinfo{volume}{7}},
Ê\bibinfo{pages}{5} (\bibinfo{year}{2004}).

\bibitem{giddings06}
\bibinfo{author}{\bibfnamefont{S.~B.} \bibnamefont{Giddings}},
Ê\bibinfo{author}{\bibfnamefont{D.}~\bibnamefont{Marolf}}, \bibnamefont{and}
Ê\bibinfo{author}{\bibfnamefont{J.~B.} \bibnamefont{Hartle}},
Ê\bibinfo{journal}{Phys. Rev. D} \textbf{\bibinfo{volume}{74}},
Ê\bibinfo{pages}{064018} (\bibinfo{year}{2006}).

\bibitem{schwinger61}
\bibinfo{author}{\bibfnamefont{J.}~\bibnamefont{Schwinger}},
Ê\bibinfo{journal}{J. Math. Phys.} \textbf{\bibinfo{volume}{2}},
Ê\bibinfo{pages}{407} (\bibinfo{year}{1961}).

\bibitem{chou85}
\bibinfo{author}{\bibfnamefont{K.}~\bibnamefont{Chou}},
Ê\bibinfo{author}{\bibfnamefont{Z.}~\bibnamefont{Su}},
Ê\bibinfo{author}{\bibfnamefont{B.}~\bibnamefont{Hao}}, \bibnamefont{and}
Ê\bibinfo{author}{\bibfnamefont{L.}~\bibnamefont{Yu}}, \bibinfo{journal}{Phys.
ÊRep.} \textbf{\bibinfo{volume}{118}}, \bibinfo{pages}{1}
Ê(\bibinfo{year}{1985}).

\bibitem{jordan86}
\bibinfo{author}{\bibfnamefont{R.~D.} \bibnamefont{Jordan}},
Ê\bibinfo{journal}{Phys. Rev. D} \textbf{\bibinfo{volume}{33}},
Ê\bibinfo{pages}{444} (\bibinfo{year}{1986}).

\bibitem{calzetta87}
\bibinfo{author}{\bibfnamefont{E.}~\bibnamefont{Calzetta}} \bibnamefont{and}
Ê\bibinfo{author}{\bibfnamefont{B.~L.} \bibnamefont{Hu}},
Ê\bibinfo{journal}{Phys. Rev. D} \textbf{\bibinfo{volume}{35}},
Ê\bibinfo{pages}{495} (\bibinfo{year}{1987}).

\bibitem{campos94}
\bibinfo{author}{\bibfnamefont{A.}~\bibnamefont{Campos}} \bibnamefont{and}
Ê\bibinfo{author}{\bibfnamefont{E.}~\bibnamefont{Verdaguer}},
Ê\bibinfo{journal}{Phys. Rev. D} \textbf{\bibinfo{volume}{49}},
Ê\bibinfo{pages}{1861} (\bibinfo{year}{1994}).

\bibitem{weinberg05}
\bibinfo{author}{\bibfnamefont{S.}~\bibnamefont{Weinberg}},
Ê\bibinfo{journal}{Phys. Rev. D} \textbf{\bibinfo{volume}{72}},
Ê\bibinfo{pages}{043514} (\bibinfo{year}{2005}).

\bibitem{calzetta94}
\bibinfo{author}{\bibfnamefont{E.}~\bibnamefont{Calzetta}} \bibnamefont{and}
Ê\bibinfo{author}{\bibfnamefont{B.~L.} \bibnamefont{Hu}},
Ê\bibinfo{journal}{Phys. Rev. D} \textbf{\bibinfo{volume}{49}},
Ê\bibinfo{pages}{6636} (\bibinfo{year}{1994}).

\bibitem{martin99a}
\bibinfo{author}{\bibfnamefont{R.}~\bibnamefont{Mart\'{\i }n}}
Ê\bibnamefont{and}
Ê\bibinfo{author}{\bibfnamefont{E.}~\bibnamefont{Verdaguer}},
Ê\bibinfo{journal}{Phys. Lett. B} \textbf{\bibinfo{volume}{465}},
Ê\bibinfo{pages}{113} (\bibinfo{year}{1999}).

\bibitem{martin99b}
\bibinfo{author}{\bibfnamefont{R.}~\bibnamefont{Mart\'{\i }n}}
Ê\bibnamefont{and}
Ê\bibinfo{author}{\bibfnamefont{E.}~\bibnamefont{Verdaguer}},
Ê\bibinfo{journal}{Phys. Rev. D} \textbf{\bibinfo{volume}{60}},
Ê\bibinfo{pages}{084008} (\bibinfo{year}{1999}).

\bibitem{hu03a}
\bibinfo{author}{\bibfnamefont{B.~L.} \bibnamefont{Hu}} \bibnamefont{and}
Ê\bibinfo{author}{\bibfnamefont{E.}~\bibnamefont{Verdaguer}},
Ê\bibinfo{journal}{Class. Quant. Grav.} \textbf{\bibinfo{volume}{20}},
Ê\bibinfo{pages}{R1} (\bibinfo{year}{2003}).

\bibitem{hu08}
\bibinfo{author}{\bibfnamefont{B.~L.} \bibnamefont{Hu}} \bibnamefont{and}
Ê\bibinfo{author}{\bibfnamefont{E.}~\bibnamefont{Verdaguer}},
Ê\bibinfo{journal}{Living Rev. Rel.} \textbf{\bibinfo{volume}{11}},
Ê\bibinfo{pages}{3} (\bibinfo{year}{2008}).

\bibitem{hu04b}
\bibinfo{author}{\bibfnamefont{B.~L.} \bibnamefont{Hu}},
Ê\bibinfo{author}{\bibfnamefont{A.}~\bibnamefont{Roura}}, \bibnamefont{and}
Ê\bibinfo{author}{\bibfnamefont{E.}~\bibnamefont{Verdaguer}},
Ê\bibinfo{journal}{Phys. Rev. D} \textbf{\bibinfo{volume}{70}},
Ê\bibinfo{pages}{044002} (\bibinfo{year}{2004}).

\bibitem{tomboulis77}
\bibinfo{author}{\bibfnamefont{E.}~\bibnamefont{Tomboulis}},
Ê\bibinfo{journal}{Phys. Lett. B} \textbf{\bibinfo{volume}{70}},
Ê\bibinfo{pages}{361} (\bibinfo{year}{1977}).

\bibitem{stewart90}
\bibinfo{author}{\bibfnamefont{J.~M.} \bibnamefont{Stewart}},
Ê\bibinfo{journal}{Class. Quant. Grav.} \textbf{\bibinfo{volume}{7}},
Ê\bibinfo{pages}{1169} (\bibinfo{year}{1990}).

\bibitem{sloth06}
\bibinfo{author}{\bibfnamefont{M.~S.} \bibnamefont{Sloth}},
Ê\bibinfo{journal}{Nucl. Phys. B} \textbf{\bibinfo{volume}{748}},
Ê\bibinfo{pages}{149} (\bibinfo{year}{2006}).

\bibitem{sloth07}
\bibinfo{author}{\bibfnamefont{M.~S.} \bibnamefont{Sloth}},
Ê\bibinfo{journal}{Nucl. Phys. B} \textbf{\bibinfo{volume}{775}},
Ê\bibinfo{pages}{78} (\bibinfo{year}{2007}).

\bibitem{seery07}
\bibinfo{author}{\bibfnamefont{D.}~\bibnamefont{Seery}},
Ê\bibinfo{journal}{JCAP} \textbf{\bibinfo{volume}{11}}, \bibinfo{pages}{025}
Ê(\bibinfo{year}{2007}).

\bibitem{roura08}
\bibinfo{author}{\bibfnamefont{A.}~\bibnamefont{Roura}} \bibnamefont{and}
Ê\bibinfo{author}{\bibfnamefont{E.}~\bibnamefont{Verdaguer}},
Ê\bibinfo{journal}{Phys. Rev. D} \textbf{\bibinfo{volume}{78}},
Ê\bibinfo{pages}{064010} (\bibinfo{year}{2008}).

\bibitem{maldacena03b}
\bibinfo{author}{\bibfnamefont{J.}~\bibnamefont{Maldacena}},
Ê\bibinfo{journal}{JHEP} \textbf{\bibinfo{volume}{05}}, \bibinfo{pages}{013}
Ê(\bibinfo{year}{2003}).

\bibitem{seery08}
\bibinfo{author}{\bibfnamefont{D.}~\bibnamefont{Seery}},
Ê\bibinfo{journal}{JCAP} \textbf{\bibinfo{volume}{02}}, \bibinfo{pages}{006}
Ê(\bibinfo{year}{2008}).

\bibitem{adshead09b}
\bibinfo{author}{\bibfnamefont{P.}~\bibnamefont{Adshead}},
Ê\bibinfo{author}{\bibfnamefont{R.}~\bibnamefont{Easther}}, \bibnamefont{and}
Ê\bibinfo{author}{\bibfnamefont{E.~A.} \bibnamefont{Lim}},
Ê\bibinfo{journal}{Phys. Rev. D} \textbf{\bibinfo{volume}{80}},
Ê\bibinfo{pages}{083521} (\bibinfo{year}{2009}).

\bibitem{urakawa08b}
\bibinfo{author}{\bibfnamefont{Y.}~\bibnamefont{Urakawa}} \bibnamefont{and}
Ê\bibinfo{author}{\bibfnamefont{K.}~\bibnamefont{Maeda}},
Ê\bibinfo{journal}{Phys. Rev. D} \textbf{\bibinfo{volume}{78}},
Ê\bibinfo{pages}{064004} (\bibinfo{year}{2008}).

\bibitem{anderson00}
\bibinfo{author}{\bibfnamefont{P.~R.} \bibnamefont{Anderson}},
Ê\bibinfo{author}{\bibfnamefont{W.}~\bibnamefont{Eaker}},
Ê\bibinfo{author}{\bibfnamefont{S.}~\bibnamefont{Habib}},
Ê\bibinfo{author}{\bibfnamefont{C.}~\bibnamefont{Molina-Par\'{\i}s}},
Ê\bibnamefont{and} \bibinfo{author}{\bibfnamefont{E.}~\bibnamefont{Mottola}},
Ê\bibinfo{journal}{Phys. Rev. D} \textbf{\bibinfo{volume}{62}},
Ê\bibinfo{pages}{124019} (\bibinfo{year}{2000}).

\bibitem{isaacson91}
\bibinfo{author}{\bibfnamefont{J.~A.} \bibnamefont{Isaacson}} \bibnamefont{and}
Ê\bibinfo{author}{\bibfnamefont{B.}~\bibnamefont{Rogers}},
Ê\bibinfo{journal}{Nucl. Phys. B} \textbf{\bibinfo{volume}{364}},
Ê\bibinfo{pages}{381} (\bibinfo{year}{1991}).

\bibitem{rogers92}
\bibinfo{author}{\bibfnamefont{B.}~\bibnamefont{Rogers}} \bibnamefont{and}
Ê\bibinfo{author}{\bibfnamefont{J.~A.} \bibnamefont{Isaacson}},
Ê\bibinfo{journal}{Nucl. Phys. B} \textbf{\bibinfo{volume}{368}},
Ê\bibinfo{pages}{415} (\bibinfo{year}{1992}).

\bibitem{busch92}
\bibinfo{author}{\bibfnamefont{C.}~\bibnamefont{Busch}} (\bibinfo{year}{1992}),
Ê\eprint{DESY-92-131, ITP-UH-9-92; arXiv:0803.3204 [gr-qc]}.

\bibitem{perez-nadal08a}
\bibinfo{author}{\bibfnamefont{G.}~\bibnamefont{P\'erez-Nadal}},
Ê\bibinfo{author}{\bibfnamefont{A.}~\bibnamefont{Roura}}, \bibnamefont{and}
Ê\bibinfo{author}{\bibfnamefont{E.}~\bibnamefont{Verdaguer}},
Ê\bibinfo{journal}{Phys. Rev. D} \textbf{\bibinfo{volume}{77}},
Ê\bibinfo{pages}{124033} (\bibinfo{year}{2008}).

\bibitem{perez-nadal08b}
\bibinfo{author}{\bibfnamefont{G.}~\bibnamefont{P\'erez-Nadal}},
Ê\bibinfo{author}{\bibfnamefont{A.}~\bibnamefont{Roura}}, \bibnamefont{and}
Ê\bibinfo{author}{\bibfnamefont{E.}~\bibnamefont{Verdaguer}},
Ê\bibinfo{journal}{Class. Quant. Grav.} \textbf{\bibinfo{volume}{25}},
Ê\bibinfo{pages}{154013} (\bibinfo{year}{2008}).

\bibitem{chernikov68}
\bibinfo{author}{\bibfnamefont{N.~A.} \bibnamefont{Chernikov}}
Ê\bibnamefont{and} \bibinfo{author}{\bibfnamefont{E.~A.}
Ê\bibnamefont{Tagirov}}, \bibinfo{journal}{Ann. Inst. Henri Poincar\'e A}
Ê\textbf{\bibinfo{volume}{9}}, \bibinfo{pages}{109} (\bibinfo{year}{1968}).

\end{thebibliography}

\end{document}